\documentclass[11pt,twoside]{article}

\usepackage[T1]{fontenc}
\usepackage[utf8]{inputenc}
\usepackage{amsmath,amssymb}
\usepackage{graphicx}
\usepackage{rotating}          % sidewaystable for Table 4
\usepackage{booktabs}
\usepackage{array}
\usepackage{caption}
\usepackage{natbib}
\usepackage[a4paper,margin=25mm]{geometry}
\usepackage{xcolor}
\usepackage[section]{placeins}  
\usepackage{hyperref}
\usepackage[percent]{overpic}
\usepackage{float}

\hypersetup{colorlinks=true,linkcolor=black,citecolor=black,urlcolor=blue}

\bibpunct{(}{)}{;}{a}{}{,}

\newcommand{\HII}{H\,\textsc{ii}}
\newcommand{\HI}{H\,\textsc{i}}
\newcommand{\NH}{N_{\mathrm{H\,\textsc{i}+H_2}}}
\newcommand{\Av}{A_V}
\newcommand{\Tdust}{T_{\mathrm{dust}}}
\newcommand{\Tgas}{T_{\mathrm{gas}}}
\newcommand{\Tex}{T_{\mathrm{ex}}}
\newcommand{\Trot}{T_{\mathrm{rot}}}
\newcommand{\Tmb}{T_{\mathrm{mb}}}
\newcommand{\kms}{\mathrm{km\,s^{-1}}}
\newcommand{\percc}{\mathrm{cm^{-3}}}
\newcommand{\persc}{\mathrm{cm^{-2}}}
\newcommand{\mg}{^{\mathrm{m}}}         
\newcommand{\code}[1]{\texttt{#1}}

\begin{document}

\title{\bfseries Gas-phase and Surface Chemistry in the Massive\\
       Star-Forming Region RCW\,120}

\author{K.~V.~Plakitina$^1$\thanks{E-mail: \texttt{plakitina.kv@inasan.ru}},
        \and M.~S.~Kirsanova$^1$, \and D.~S.~Wiebe$^1$, \and O.~V.~Kochina$^1$}

\date{$^1$Institute of Astronomy of the Russian Academy of Sciences,
      Moscow, 119017 Russia\\[2pt]
      \small Received February 9, 2025; revised April 1, 2025;
      accepted May 5, 2025}

\maketitle

%======================================================================
\begin{abstract}
\noindent
Molecules in the interstellar medium form both in the gas phase and on
dust grains. The chemical pathways of molecular formation are not yet
understood in detail, so the question of which pathway dominates for a
particular molecule remains open. We analysed broadband emission
spectra of a dense molecular clump in RCW\,120, obtained with the APEX
telescope in the 200--260\,GHz range, in order to investigate molecular
formation pathways in regions of massive star formation at an early
evolutionary stage. We examined the correlations between the molecular
column densities derived under the LTE assumption. An excess of
methanol was found in the southern part of the dense clump relative to
its northern part, while the abundances of other molecules, such as
CH$_3$CN and CH$_3$CCH, remain comparable. The methanol abundance is
also elevated relative to that of other oxygen-bearing molecules, such
as OCS and SO. To identify possible causes of the enhanced methanol
abundance in the southern part of the clump, we carried out simulations
with the astrochemical model \code{Presta} in a two-phase
approximation, accounting for chemical processes both in the gas phase
and in the mantles of dust grains. The modelling shows that the enhanced
gas-phase methanol abundance may be due to photodesorption from icy
mantles. At $\Av$ values between $4\mg$ and $6\mg$, methanol desorbs efficiently from the ice mantles of dust grains upon
interaction with photons, but is not yet destroyed by UV radiation in
the gas phase. A strong linear correlation between molecular column
densities indicates that the molecules form in the same phase --- either in the gas phase or on dust. Their integrated intensity maps may
nevertheless differ, as is the case for CCH and CH$_3$CN. If two
molecules form in different phases --- one in the gas phase and the other
in dust mantles --- no correlation is observed, as for CCH and
CH$_3$OH. The weak correlation between methanol and the oxygen-bearing
molecules that form on dust suggests that only the upper part of the
dust mantles, rich in CO ice, is destroyed in the southern part of the
clump.
\end{abstract}

\noindent\textbf{DOI:} 10.1134/S1990341325600073


%======================================================================
\section{INTRODUCTION}
\label{sec:intro}

Surveys of star-forming regions in the millimetre range with ultra-high
spectral resolution ($R\geq10^7$) have become routine
\citep{Pety2017}, including those covering the Galactic plane
\citep{Rathborne2016}. As a result, the need for effective
interpretation of the resulting rich observational datasets has become
increasingly pressing. One way to address this problem is to develop
statistical methods that would allow the physical parameters of
star-forming regions to be determined efficiently, and the processes
occurring in them to be studied, from the observed brightnesses of
molecular lines and the correlations between them
\citep{Orkisz2017,Gratier2021}. The application of these methods is
complicated by the fact that the brightness distribution of molecular
emission is determined not only by the spatial distribution of matter
and by the excitation conditions, but also by the complexity of the
chemical processes operating under diverse physical conditions.

In this work we consider the possibility of investigating astrochemical
pathways of molecule formation by means of statistical analysis.
Numerous observations and models show, for example, that small
hydrocarbons such as CCH form in the gas phase, and that this process
can proceed both from simple compounds to complex ones
\citep{Murga2020,Murga2023} and vice versa
\citep{Pety2005,Awad2022}. Other simple molecules, such as SO and OCS
\citep{Ferrante2008}, can also form on dust; dust is especially
important as a catalyst for the formation of complex organic
molecules---methanol and even more complex compounds
\citep{Watanabe2002,Fuchs2009,Punanova2022}. On the other hand, some
complex molecules, such as CH$_3$CCH and other long carbon chains, form
in the gas phase \citep{Agundez2021,Byrne2023}. Finally, many
molecules, such as water, have several chemical formation channels
\citep[see, for example,][]{vanDishoeck2021}, and for them the question
of which channel dominates under given physical conditions is
particularly relevant.

The observational data we use cover a dense molecular clump with young
stellar objects (YSOs) embedded in it: RCW\,120\,S1, RCW\,120\,S2,
RCW\,120\,S9, RCW\,120\,S10 and RCW\,120\,S39 (hereafter, for
brevity, we omit the designation RCW\,120). In this region the
molecular abundances are set by the simultaneous action of several
factors. The clump is located at the edge of the \HII{} region of
RCW\,120, so the molecular abundances near the ionization front are
affected by the ultraviolet (UV) radiation of the massive star. The
molecular abundances and the brightness of the molecular lines may also
depend on the presence of YSOs in the dense molecular gas; we test this
possibility here as well.

\begin{figure}[tbp]
\centering
\begin{overpic}[width=\textwidth]{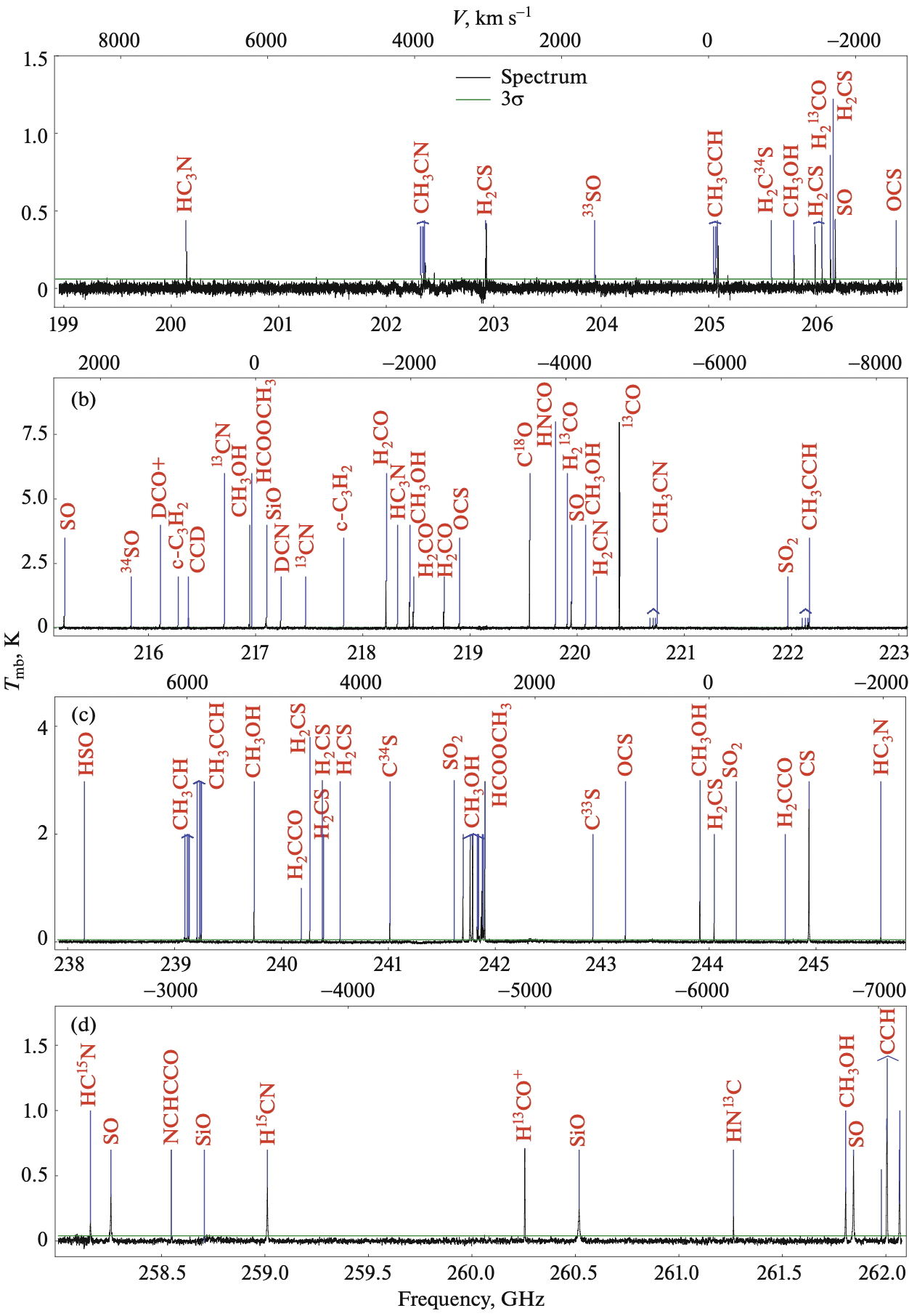}
  \put(5, 93){\small \textbf{(a)}}  
\end{overpic}
% \includegraphics[width=\textwidth]{figs/full_spectra.png}
% \figplaceholder{0.85\textwidth}{100mm}
\caption{Broadband spectrum from the nFLASH230 receiver towards
YSO S2. The molecules whose lines are identified are indicated. The
solid green line shows the $3\sigma$ level, where $\sigma = 19$\,mK
(a), 21\,mK (b), 12\,mK (c) and (d).}
\label{fig:spectra}
\end{figure}

Our study is based on observational data obtained with the APEX
telescope \citep{Gusten2006} using the nFLASH230 receiver
\citep{Belitsky2018} and described in our previous paper
\citep{Plakitina2024}, where we presented results for the 20 brightest
molecular emission lines. The broadband spectra obtained with
nFLASH230 are shown in Fig.~\ref{fig:spectra}. In this paper we
complete the analysis of these observational data.

In Section~\ref{sec:methods} we describe the observational data, the
method used to estimate the molecular column densities, and the model
with which we compare the observations. In
Section~\ref{sec:results} we perform a correlation analysis of the
column densities and, using the model, search for the causes of the
observed pattern. In Section~\ref{sec:discussion} we discuss the
destruction of dust grain mantles and how it affects the abundance of
molecules in the gas phase.

%======================================================================
\section{METHODS}
\label{sec:methods}

\subsection{Identification of Spectral Lines}
\label{sec:lines}

The spectral emission lines of molecules in the observed frequency
bands were identified with the \code{WEEDS} extension of the
\code{CLASS} software package \citep{Maret2011}, using the Cologne
Database for Molecular Spectroscopy
\citep[CDMS,][]{Muller2001}\footnote{\url{https://cdms.astro.uni-koeln.de}}.
As in \citet{Plakitina2024}, we consider a line to be detected if its
peak brightness exceeds the threshold value $3\sigma$, where $\sigma$
is the standard deviation of the noise in the spectrum. The lines
identified towards the five YSOs are shown in
Fig.~\ref{fig:linespanels1},~\ref{fig:linespanels2} and their spectroscopic parameters are
listed in Table~\ref{tab:spectro}.

\begin{table}[H]
\centering
\caption{Spectroscopic parameters of the detected spectral lines taken
from the CDMS database: molecule, transition quantum numbers, frequency
$\nu$ and upper level energy $E_u$.}
\label{tab:spectro}
\scriptsize
\setlength{\tabcolsep}{4pt}
\begin{tabular}{llrr@{\hspace{3mm}}|@{\hspace{3mm}}llrr}
\toprule
Molecule & Transition & $\nu$, MHz & $E_u$, K &
Molecule & Transition & $\nu$, MHz & $E_u$, K \\
\midrule
HC$_3$N        & $J=22-21$                          & 200\,135.392 & 110.5 & SO$_2$        & $J_{K_a,K_c}=11_{1,11}-10_{0,10}$ & 221\,965.220 & 60.4 \\
H$_2$CS        & $J_{K_a,K_c}=6_{1,6}-5_{1,5}$      & 202\,924.054 & 47.3  & HSO           & $N_{K_a,K_c}=6_{2,5}-5_{2,4}$,    & 238\,156.771 & 93.4 \\
$^{33}$SO      & $J_{K_a,K_c}=5_{4,6}-4_{3,5}$      & 203\,939.255 & 38.3  &               & $J=13/2-11/2$, $F=6-5$            &              &      \\
H$_2$CS        & $J_{K_a,K_c}=6_{0,6}-5_{0,5}$      & 205\,987.858 & 34.6  & H$_2$CCO      & $J_{K_a,K_c}=12_{1,12}-11_{1,11}$ & 240\,185.794 & 88.0 \\
H$_2$CS        & $J_{K_a,K_c}=6_{3,4}-5_{3,3}$      & 206\,052.602 & 153.1 & H$_2$CS       & $J_{K_a,K_c}=7_{0,7}-6_{0,6}$     & 240\,266.872 & 46.1 \\
H$_2^{13}$CO   & $J_{K_a,K_c}=3_{1,3}-2_{1,2}$      & 206\,131.626 & 31.6  & H$_2$CS       & $J_{K_a,K_c}=7_{2,6}-6_{2,5}$     & 240\,382.051 & 98.8 \\
H$_2$CS        & $J_{K_a,K_c}=6_{2,4}-5_{2,3}$      & 206\,158.602 & 87.3  & H$_2$CS       & $J_{K_a,K_c}=7_{3,5}-6_{3,4}$     & 240\,393.037 & 164.6 \\
SO             & $J_K=5_4-4_3$                      & 206\,176.005 & 38.6  & H$_2$CS       & $J_{K_a,K_c}=7_{3,4}-6_{3,3}$     & 240\,393.762 & 164.6 \\
OCS            & $J=17-16$                          & 206\,745.156 & 89.3  & H$_2$CS       & $J_{K_a,K_c}=7_{2,5}-6_{2,4}$     & 240\,549.066 & 98.8 \\
SO             & $J_K=5_5-4_4$                      & 215\,220.653 & 44.1  & SO$_2$        & $J_{K_a,K_c}=5_{2,4}-4_{1,3}$     & 241\,615.797 & 23.6 \\
$^{34}$SO      & $J_K=5_6-4_5$                      & 215\,839.920 & 34.4  & C$^{33}$S     & $J_K=5_0-4_0$                     & 242\,913.610 & 35.0 \\
c-C$_3$H$_2$   & $J_{K_a,K_c}=3_{3,0}-2_{2,1}$      & 216\,278.756 & 19.5  & OCS           & $J=20-19$                         & 243\,218.036 & 122.6 \\
CCD            & $J_{K_a,K_c}=3_{4,5}-2_{3,4}$      & 216\,372.837 & 20.8  & H$_2$CS       & $J_{K_a,K_c}=7_{1,6}-6_{1,5}$     & 244\,048.504 & 60.0 \\
$^{13}$CN      & $N=2-1$, $J=3/2-3/2$,              & 216\,710.144 & 15.6  & SO$_2$        & $J_{K_a,K_c}=14_{0,14}-13_{1,13}$ & 244\,254.218 & 93.9 \\
               & $F_1=2-1$, $F=1-2$                 &              &       & H$_2$CCO      & $J_{K_a,K_c}=12_{1,11}-11_{1,10}$ & 244\,712.269 & 89.4 \\
HCOOCH$_3$     & $J_{K_a,K_c}=20_{0,20}-19_{1,19}$, $A$ & 216\,964.157 & 111.5 & HC$_3$N   & $J=27-26$                         & 245\,606.320 & 165.0 \\
$^{13}$CN      & $N=2-1$, $J=5/2-3/2$,              & 217\,467.150 & 15.7  & HC$^{15}$N    & $J=3-2$                           & 258\,156.996 & 24.8 \\
               & $F_1=3-2$, $F=4-3$                 &              &       & HN$^{13}$C    & $J=3-2$                           & 261\,263.513 & 25.1 \\
HC$_3$N        & $J=24-23$                          & 218\,324.723 & 131.0 & SO            & $J_K=6_7-5_6$                     & 261\,843.721 & 47.6 \\
H$_2$CO        & $J_{K_a,K_c}=3_{2,1}-2_{2,0}$      & 218\,760.066 & 68.1  & CCH           & $N_J=3_{5/2}-2_{3/2}$, $F=3-2$    & 262\,064.986 & 25.2 \\
OCS            & $J=18-17$                          & 218\,903.356 & 99.8  & CCH           & $N_J=3_{5/2}-2_{3/2}$, $F=2-1$    & 262\,067.469 & 25.2 \\
H$_2^{13}$CO   & $J_{K_a,K_c}=3_{1,2}-2_{1,1}$      & 219\,908.525 & 32.9  &               &                                   &              &      \\
H$_2$CN        & $J_{K_a,K_c}=3_{0,3}-2_{0,2}$,     & 220\,178.897 & 21.1  &               &                                   &              &      \\
               & $F=5/2-5/2$, $n=18-10$             &              &       &               &                                   &              &      \\
\bottomrule
\end{tabular}
\end{table}

\begin{figure}[H]
\centering
\includegraphics[width=\textwidth]{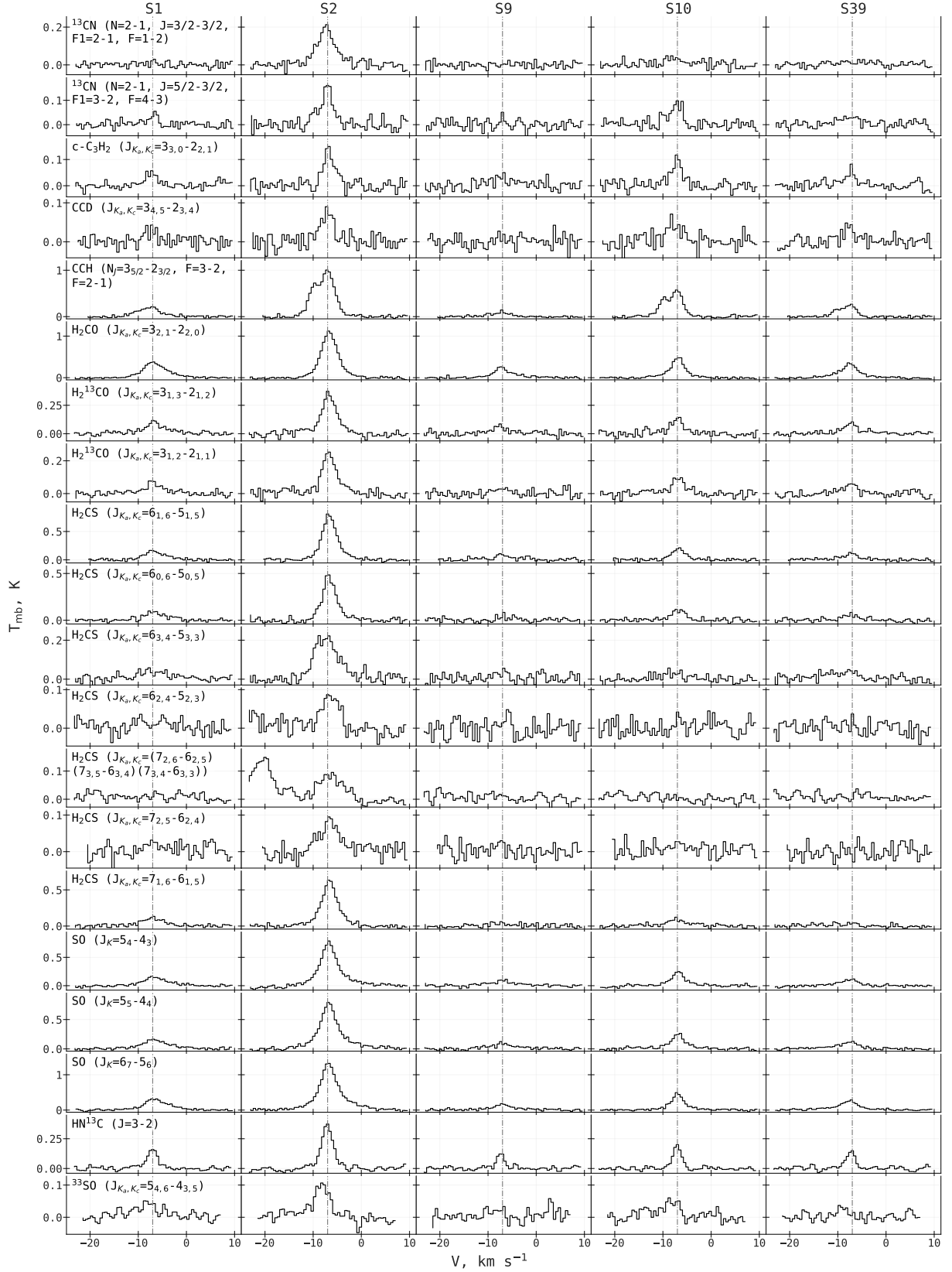}
% \figplaceholder{0.95\textwidth}{120mm}
\caption{Spectral lines detected towards YSOs S1, S2, S9, S10 and S39
in RCW\,120. The dashed gray line marks the velocity
$V_{\mathrm{LSR}} = -7\,\kms$.}

\label{fig:linespanels1}
\end{figure}

\begin{figure}[H]
\centering
\includegraphics[width=\textwidth]{figs/spectra_sources_p2.pdf}
% \figplaceholder{0.95\textwidth}{120mm}
\caption{(Cont.)}

\label{fig:linespanels2}
\end{figure}

\subsection{Estimation of Molecular Abundances}
\label{sec:abund}

The molecular column densities were estimated in the LTE
approximation, which is justified because we are studying a dense
molecular clump \citep[see, for example,][]{Kirsanova2019,%
Kirsanova2021,Kirsanova2023}. To estimate the molecular column
densities in the optically thin case, $N_{\mathrm{tot}}^{\mathrm{thin}}$,
we used the equation of \citet{Mangum2015}:
\begin{align}
N_{\mathrm{tot}}^{\mathrm{thin}}
 &= \left(\frac{3h}{8\pi^{3}S\mu^{2}R_i}\right)
    \left(\frac{Q(\Trot)}{g_u}\right)
    \frac{\exp\!\left(\dfrac{E_u}{k_B \Tex}\right)}
         {\exp\!\left(\dfrac{h\nu}{k_B \Tex}\right)-1} \nonumber\\[4pt]
 &= \frac{1}{J_\nu(\Tex)-J_\nu(T_{\mathrm{bg}})}
    \int \frac{\Tmb\, dV}{f},
\label{eq:coldens}
\end{align}
where $S$ is the line strength, $\mu$ is the dipole moment of the
molecule, $R_i$ are the relative intensities of the hyperfine
components, $\Trot$ is the rotational temperature, $Q(\Trot)$ is the
partition function over rotational transitions, $g_u$ is the
statistical weight of the upper level, $E_u$ is the upper level energy,
$k_B$ is the Boltzmann constant, $\Tex$ is the excitation temperature
of the line, and $J_\nu(T)$ is the equivalent intensity of a blackbody
at temperature $T$:
\begin{equation}
J_\nu(T) \equiv \frac{h\nu}{k_B}
  \Big/ \left[\exp\!\left(\frac{h\nu}{k_B T}\right)-1\right].
\end{equation}
In what follows we assume a beam filling factor $f = 1$ and
$T_{\mathrm{bg}} = 2.7$\,K.

For most molecules we adopted the dust temperature $\Tdust$ as $\Tex$.
In dense molecular clumps, where the state of the gas is close to LTE,
the dust temperature should be close to the gas temperature $\Tgas$ and
to the molecular excitation temperature. By using $\Tdust$ as $\Tex$ we
obtain only a lower limit on the molecular column densities, since we
have no way to estimate the optical depths of the emission lines.

For the H$_2$CS, OCS, HC$_3$N, SO and SO$_2$ molecules, which show
multiple emission lines, we constructed population diagrams from which
we estimated $\Tex$ \citep{Goldsmith1999,Kalenskii2016}.

To estimate the abundances of the molecules relative to the number of
hydrogen nuclei, $N(\HI) + 2N(\mathrm{H_2})$ (hereafter denoted $\NH$ for brevity), we divided the pixel-by-pixel maps of the molecular column densities by the values of $\NH$ obtained in
\citet{Plakitina2024}.

\subsection{Correlations of Column Densities}
\label{sec:corr}

To estimate the linear correlations between the column densities of
different pairs of molecules we used the Pearson correlation
coefficient:
\begin{equation}
p = \frac{\sum \left(x_{ji}-\bar{x}\right)\left(y_{ji}-\bar{y}\right)}
         {\sqrt{\sum\left(x_{ji}-\bar{x}\right)^{2}
                \sum\left(y_{ji}-\bar{y}\right)^{2}}},
\label{eq:pearson}
\end{equation}
where $x_{ji}$ and $y_{ji}$ are the sample elements---in our case the
molecular column densities $x$ and $y$ in the pixel $[j,i]$---and
$\bar{x}$ and $\bar{y}$ are the mean values of the molecular column
densities $x$ and $y$. Pixels in which the uncertainty in $N$ exceeded
50\% for at least one molecule of the pair were excluded from the
calculation of the correlation coefficient.

\subsection{Model}
\label{sec:model}

The astrochemical modelling was performed with the \code{Presta}
software package \citep{Kochina2013}, which is designed to follow the
chemical evolution of objects at different stages of protostellar
evolution. \code{Presta} can compute chemical evolution in three
phases: in the gas phase, on the surfaces of dust grains and in the
bulk of their icy mantles \citep{Borshcheva2022}. In this work we used
the two-phase version of the model, that is, we took into account only
reactions occurring in the gas phase and on the surfaces of dust
grains. The set of species and reactions is based on the
\code{ALCHEMIC} chemical network \citep{Semenov2011}, but with
significant additions, described in part by \citet{Wiebe2019}. The
initial abundances of the chemical species are listed in
Table~\ref{tab:initial}.

\begin{table}[H]
\centering
\caption{Initial abundances of the chemical species relative to the
number of hydrogen nuclei in the \code{Presta} model.}
\label{tab:initial}
\begin{tabular}{cc}
\toprule
Species & Initial abundance \\
\midrule
H$_2$ & $4.99995 \times 10^{-1}$ \\
S     & $8.0 \times 10^{-8}$ \\
H     & $1.0 \times 10^{-5}$ \\
Si    & $8.0 \times 10^{-9}$ \\
He    & $9.0 \times 10^{-2}$ \\
Fe    & $3.0 \times 10^{-9}$ \\
C     & $7.3 \times 10^{-5}$ \\
Mg    & $7.0 \times 10^{-9}$ \\
Cl    & $1.0 \times 10^{-9}$ \\
Na    & $2.0 \times 10^{-9}$ \\
N     & $2.14 \times 10^{-5}$ \\
P     & $2.0 \times 10^{-10}$ \\
O     & $1.76 \times 10^{-4}$ \\
\bottomrule
\end{tabular}
\end{table}

Chemical evolution was simulated over an interval of up to $10^{6}$
years under constant physical conditions. We performed several
calculations, varying the number density of hydrogen nuclei $n_{\rm H}$
in the model cloud from $10^{3}$ to $10^{5}\,\percc$, the gas
temperature $\Tgas$ from 10 to 100\,K in steps of 10\,K, and the visual
extinction $\Av$ from $3\mg$ to $6\mg$. The dust temperature was
taken to be 22\,K, based on the analysis of the IR data
\citep{Plakitina2024}.

%======================================================================
\section{RESULTS}
\label{sec:results}

\subsection{Detected Molecules}
\label{sec:detected}

Towards the emission maximum of the molecules---YSO S2---40 lines of 20
molecules, including isotopologues, were detected. They are shown in
Figs.~\ref{fig:linespanels1} and \ref{fig:linespanels2}, and listed in
Table~\ref{tab:spectro}. The
spectra were smoothed to a spectral resolution of $0.5\,\kms$, and the
resulting noise level was 12--20\,mK. Together with the results of
\citet{Plakitina2024}, 65 lines (we count a series of $J_K$ lines as a
single line) belonging to 35 molecules were detected towards YSO S2.
The most complex of them---HCOOCH$_3$ (methyl formate)---consists of
eight atoms. For the subsequent analysis the spectra were fitted with
Gaussian functions. The line parameters from which we constructed the
population diagrams are given in Tables~\ref{tab:intint} (integrated
intensities) and \ref{tab:gauss} (amplitudes, widths and radial
velocities of the Gaussians).

\begin{table}[H]
\centering
\caption{Parameters of the SO$_2$, OCS, HC$_3$N, H$_2$CS and SO lines
towards the YSOs.}
\label{tab:intint}
\scriptsize
\setlength{\tabcolsep}{4pt}
\begin{tabular}{llrrrr rrrrr}
\toprule
& \multicolumn{1}{c}{Frequency,} & & & & &
\multicolumn{5}{c}{$\int \Tmb\,dV$, mK\,$\kms$} \\
\cmidrule(l){7-11}
Molecule & \multicolumn{1}{c}{MHz} & $J_K$ & $E_u$, K &
$\log A_{ij}$ & $g_u$ & S1 & S2 & S9 & S10 & S39 \\
\midrule
SO$_2$
 & 221\,965.220 & $11_{1,11}$--$10_{0,10}$ & 60.4  & $-3.943$ & 23 & $261 \pm 2$  & $848 \pm 6$   & --            & $159 \pm 2$   & $240 \pm 1$ \\
 & 241\,615.797 & $5_{2,4}$--$4_{1,3}$     & 23.6  & $-4.073$ & 11 & $195 \pm 2$  & $465 \pm 3$   & --            & $151 \pm 2$   & -- \\
 & 244\,254.218 & $14_{0,14}$--$13_{0,13}$ & 93.9  & $-3.785$ & 29 & $157 \pm 1$  & $584 \pm 4$   & --            & --            & -- \\
\midrule
OCS
 & 206\,745.156 & 17--16 &  89.3 & $-4.592$ & 35 & $389 \pm 4$  & $1320 \pm 12$ & $209 \pm 2$   & $263 \pm 3$   & $254 \pm 2$ \\
 & 218\,903.356 & 18--17 &  99.8 & $-4.517$ & 37 & $270 \pm 3$  & $1174 \pm 10$ & $195 \pm 2$   & $190 \pm 2$   & $204 \pm 1$ \\
 & 243\,218.036 & 20--19 & 122.6 & $-4.517$ & 41 & $206 \pm 2$  & $960 \pm 9$   & --            & --            & -- \\
\midrule
HC$_3$N
 & 200\,135.392 & 22--21 & 110.5 & $-3.197$ & 35 & $194 \pm 3$  & $1631 \pm 13$ & $440 \pm 3$   & $353 \pm 4$   & $204 \pm 1$ \\
 & 218\,324.723 & 24--23 & 131.0 & $-3.083$ & 49 & $102 \pm 2$  & $1260 \pm 10$ & --            & $222 \pm 2$   & $196 \pm 1$ \\
 & 245\,606.320 & 27--26 & 165.0 & $-2.928$ & 55 & $115 \pm 3$  & $622 \pm 5$   & --            & --            & -- \\
\midrule
H$_2$CS
 & 202\,924.054 & $6_{1,6}$--$5_{1,5}$ &  47.3 & $-3.924$ & 39 & $850 \pm 7$  & $3128 \pm 38$ & $362 \pm 4$   & $731 \pm 6$   & $401 \pm 6$ \\
 & 205\,987.858 & $6_{0,6}$--$5_{0,5}$ &  34.6 & $-3.893$ & 13 & $464 \pm 4$  & $1871 \pm 23$ & $217 \pm 3$   & $416 \pm 4$   & $229 \pm 4$ \\
 & 206\,052.602 & $6_{3,4}$--$5_{3,3}$ & 153.1 & $-4.018$ & 39 & $215 \pm 2$  & $882 \pm 11$  & --            & --            & $140 \pm 3$ \\
 & 206\,158.602 & $6_{2,4}$--$5_{2,3}$ &  87.3 & $-3.943$ & 13 & --           & $359 \pm 5$   & --            & --            & -- \\
 & 240\,266.872 & $7_{0,7}$--$6_{0,6}$ &  46.1 & $-3.688$ & 15 & $276 \pm 3$  & $1460 \pm 18$ & --            & --            & $253 \pm 4$ \\
 & 240\,382.051 & $7_{2,6}$--$6_{2,5}$ &  60.0 & $-3.678$ & 45 & --           & $347 \pm 5$   & --            & --            & -- \\
 & 240\,549.066 & $7_{2,5}$--$6_{2,4}$ &  60.0 & $-3.678$ & 45 & --           & $325 \pm 5$   & --            & --            & -- \\
 & 244\,048.504 & $7_{1,6}$--$6_{1,5}$ &  98.8 & $-3.724$ & 15 & $624 \pm 5$  & $2426 \pm 30$ & --            & $302 \pm 3$   & -- \\

\midrule
SO
 & 206\,176.005 & $5_4$--$4_3$ & 38.6 & $-3.996$ &  9 & $1152 \pm 11$ & $3809 \pm 99$  & $497 \pm 6$   & $1105 \pm 16$ & $910 \pm 9$ \\
 & 215\,220.653 & $5_5$--$4_4$ & 44.1 & $-3.924$ & 11 & $3571 \pm 93$ & $435 \pm 5$    & $858 \pm 12$  & $930 \pm 50$  & $609 \pm 6$ \\

 & 219\,949.442 & $5_6$--$4_5$ & 35.0 & $-3.873$ & 13 & $3229 \pm 29$ & $8325 \pm 214$ & $1744 \pm 18$ & $2796 \pm 39$ & $2442 \pm 24$ \\
 & 258\,255.826 & $6_6$--$5_5$ & 56.5 & $-3.996$ & 13 & $744 \pm 7$   & $3100 \pm 82$  & --            & $583 \pm 9$   & $447 \pm 5$ \\
 & 261\,843.722 & $6_7$--$5_6$ & 47.6 & $-3.642$ & 15 & $1973 \pm 18$ & $6033 \pm 156$ & $855 \pm 9$   & $1596 \pm 22$ & $1340 \pm 13$ \\
\bottomrule
\end{tabular}
\end{table}

\begin{table}[H]
\centering
\rotatebox{90}{%          % или {-90} для обратного направления
\begin{minipage}{\textheight}
\centering
\caption{Parameters of the Gaussian fits to the SO$_2$, OCS, HC$_3$N,
H$_2$CS and SO lines towards the YSOs. The main-beam brightness
temperature ($\Tmb$), the full width at half maximum ($FWHM$) and the
velocity of the emission maximum ($V_{\mathrm{peak}}$) are given.}
\label{tab:gauss}
\scriptsize
\setlength{\tabcolsep}{3pt}
\begin{tabular}{ll rrrrr rrrrr rrrrr}
\toprule
& Frequency, & \multicolumn{5}{c}{$\Tmb$, mK}
& \multicolumn{5}{c}{$FWHM$, $\kms$}
& \multicolumn{5}{c}{$V_{\mathrm{peak}}$, $\kms$} \\
\cmidrule(lr){3-7}\cmidrule(lr){8-12}\cmidrule(l){13-17}
Molecule & MHz & S1 & S2 & S9 & S10 & S39
              & S1 & S2 & S9 & S10 & S39
              & S1 & S2 & S9 & S10 & S39 \\
\midrule
SO$_2$
 & 221\,965.220 & $38\pm13$ & $119\pm19$ & -- & $49\pm16$ & $29\pm13$ & $7.1\pm0.9$ & $6.7\pm0.4$ & -- & $3.0\pm0.6$ & $8.2\pm1.4$ & $-6.9\pm0.5$ & $-6.5\pm0.2$ & -- & $-7.1\pm0.6$ & $-8.3\pm0.7$ \\
 & 241\,615.797 & $28\pm9$  & $65\pm22$  & -- & $47\pm14$ & --        & $7.1\pm1.2$ & $6.7\pm0.6$ & -- & $3.0\pm0.4$ & --          & $-6.9\pm0.6$ & $-6.5\pm0.3$ & -- & $-7.1\pm0.4$ & -- \\
 & 244\,254.218 & $21\pm7$  & $82\pm21$  & -- & --        & --        & $7.1\pm1.9$ & $6.7\pm0.6$ & -- & --          & --          & $-6.9\pm0.9$ & $-6.5\pm0.3$ & -- & --           & -- \\
\midrule
OCS
 & 206\,745.156 & $74\pm14$ & $258\pm18$ & -- & $59\pm15$ & $44\pm15$ & $5.0\pm0.4$ & $4.8\pm0.2$ & -- & $4.2\pm0.6$ & $5.4\pm0.7$ & $-6.8\pm0.2$ & $-6.7\pm0.1$ & -- & $-6.6\pm0.3$ & $-8.1\pm0.4$ \\
 & 218\,903.356 & $51\pm17$ & $230\pm19$ & -- & --        & --        & $5.0\pm0.4$ & $4.8\pm0.2$ & -- & --          & --          & $-6.8\pm0.2$ & $-6.7\pm0.1$ & -- & --           & -- \\
 & 243\,218.036 & $39\pm13$ & $188\pm21$ & -- & --        & --        & $5.0\pm0.7$ & $4.8\pm0.2$ & -- & --          & --          & $-6.8\pm0.4$ & $-6.7\pm0.1$ & -- & --           & -- \\
\midrule
HC$_3$N
 & 200\,135.392 & $57\pm17$ & $329\pm22$ & -- & $76\pm22$ & -- & $5.2\pm0.7$ & $4.7\pm0.2$ & -- & $4.4\pm0.5$ & -- & $-7.2\pm0.4$ & $-6.9\pm0.1$ & -- & $-7.2\pm0.3$ & -- \\
 & 218\,324.723 & $25\pm10$ & $254\pm20$ & -- & $48\pm17$ & -- & $5.2\pm1.2$ & $4.7\pm0.1$ & -- & $4.4\pm0.7$ & -- & $-7.2\pm0.6$ & $-6.9\pm0.1$ & -- & $-7.2\pm0.4$ & -- \\
 & 245\,606.320 & $36\pm14$ & $125\pm22$ & -- & --        & -- & $5.2\pm1.5$ & $4.7\pm0.4$ & -- & --          & -- & $-7.2\pm0.8$ & $-6.9\pm0.2$ & -- & --           & -- \\
\midrule
H$_2$CS
 & 202\,924.054 & $146\pm17$ & $761\pm23$ & $60\pm21$ & $195\pm21$ & $101\pm18$ & $5.5\pm0.3$ & $3.9\pm0.1$ & $5.7\pm1.1$ & $3.5\pm0.2$ & $3.7\pm0.4$ & $-7.0\pm0.1$ & $-6.9\pm0.1$ & $-6.8\pm0.6$ & $-7.1\pm0.1$ & $-7.8\pm0.2$ \\
 & 205\,987.858 & $80\pm11$  & $455\pm23$ & $36\pm14$ & $111\pm17$ & $57\pm16$  & $5.5\pm0.4$ & $3.9\pm0.1$ & $5.7\pm1.1$ & $3.5\pm0.3$ & $3.7\pm0.5$ & $-7.0\pm0.2$ & $-6.9\pm0.1$ & $-6.8\pm0.6$ & $-7.1\pm0.1$ & $-7.8\pm0.3$ \\
 & 206\,052.602 & $37\pm13$  & $215\pm20$ & --        & --         & $35\pm12$  & $5.5\pm1.0$ & $3.9\pm0.2$ & --          & --          & $3.7\pm1.2$ & $-7.0\pm0.5$ & $-6.9\pm0.1$ & --           & --           & $-7.8\pm0.6$ \\
 & 206\,158.602 & --         & $87\pm16$  & --        & --         & --         & --          & $3.9\pm0.4$ & --          & --          & --          & --           & $-6.9\pm0.2$ & --           & --           & -- \\
 & 240\,266.872 & $47\pm17$  & $355\pm22$ & --        & --         & $65\pm18$  & $5.5\pm0.8$ & $3.9\pm0.1$ & --          & --          & $3.7\pm0.4$ & $-7.0\pm0.4$ & $-6.9\pm0.1$ & --           & --           & $-7.8\pm0.2$ \\
 & 240\,382.051 & --         & $85\pm19$  & --        & --         & --         & --          & $3.9\pm1.1$ & --          & --          & --          & --           & $-6.9\pm0.6$ & --           & --           & -- \\
 & 240\,549.066 & --         & $79\pm19$  & --        & --         & --         & --          & $3.9\pm0.5$ & --          & --          & --          & --           & $-6.9\pm0.2$ & --           & --           & -- \\
 & 244\,048.504 & $107\pm18$ & $590\pm22$ & --        & $81\pm19$  & --         & $5.5\pm0.4$ & $3.9\pm0.1$ & --          & $3.5\pm0.5$ & --          & $-7.0\pm0.2$ & $-6.9\pm0.1$ & --           & $-7.1\pm0.3$ & -- \\
\midrule
SO
 & 206\,176.005 & $169\pm15$ & $761\pm25$  & $83\pm18$  & $301\pm24$ & $162\pm17$ & $6.4\pm0.2$ & $4.7\pm0.1$ & $5.6\pm0.6$ & $3.5\pm0.1$ & $5.3\pm0.2$ & $-6.6\pm0.1$  & $-6.9\pm0.1$ & $-7.0\pm0.3$ & $-7.1\pm0.1$ & $-7.8\pm0.1$ \\
 & 215\,220.653 & $144\pm18$ & $714\pm17$  & $73\pm17$  & $234\pm21$ & $108\pm32$ & $6.4\pm0.3$ & $4.7\pm0.1$ & $5.6\pm0.6$ & $3.5\pm0.2$ & $5.3\pm0.4$ & $-6.6\pm0.13$ & $-6.9\pm0.1$ & $-7.0\pm0.3$ & $-7.1\pm0.1$ & $-7.8\pm0.2$ \\
 & 219\,949.442 & $474\pm20$ & $1663\pm23$ & $291\pm19$ & $762\pm20$ & $435\pm21$ & $6.4\pm0.2$ & $4.7\pm0.1$ & $5.6\pm0.2$ & $3.5\pm0.1$ & $5.3\pm0.2$ & $-6.6\pm0.1$  & $-6.9\pm0.1$ & $-7.0\pm0.1$ & $-7.1\pm0.1$ & $-7.8\pm0.1$ \\
 & 258\,255.826 & $109\pm17$ & $619\pm20$  & --         & $159\pm19$ & $80\pm19$  & $6.4\pm0.4$ & $4.7\pm0.1$ & --          & $3.5\pm0.2$ & $5.3\pm0.5$ & $-6.6\pm0.2$  & $-6.9\pm0.1$ & --           & $-7.1\pm0.1$ & $-7.8\pm0.3$ \\
 & 261\,843.722 & $290\pm19$ & $1206\pm20$ & $143\pm18$ & $435\pm17$ & $239\pm21$ & $6.4\pm0.2$ & $4.7\pm0.1$ & $5.6\pm0.4$ & $3.5\pm0.1$ & $5.3\pm0.2$ & $-6.6\pm0.1$  & $-6.9\pm0.1$ & $-7.0\pm0.2$ & $-7.1\pm0.1$ & $-7.8\pm0.1$ \\
\bottomrule
\end{tabular}
\end{minipage}}
\end{table}

\subsection{Molecular Line Emission Maps}
\label{sec:maps}

In \citet{Plakitina2024} we presented integrated intensity maps of the
twenty molecules whose emission towards the observed region is
brightest. Here, in Figs.~\ref{fig:moment0_p1} and \ref{fig:moment0_p2}, we present
integrated
intensity maps for the remaining molecules detected in this region,
which have weaker lines. Like the molecules considered in the previous
work, these can be divided into three groups.

\begin{figure}[p]
\centering
\includegraphics[width=\textwidth]{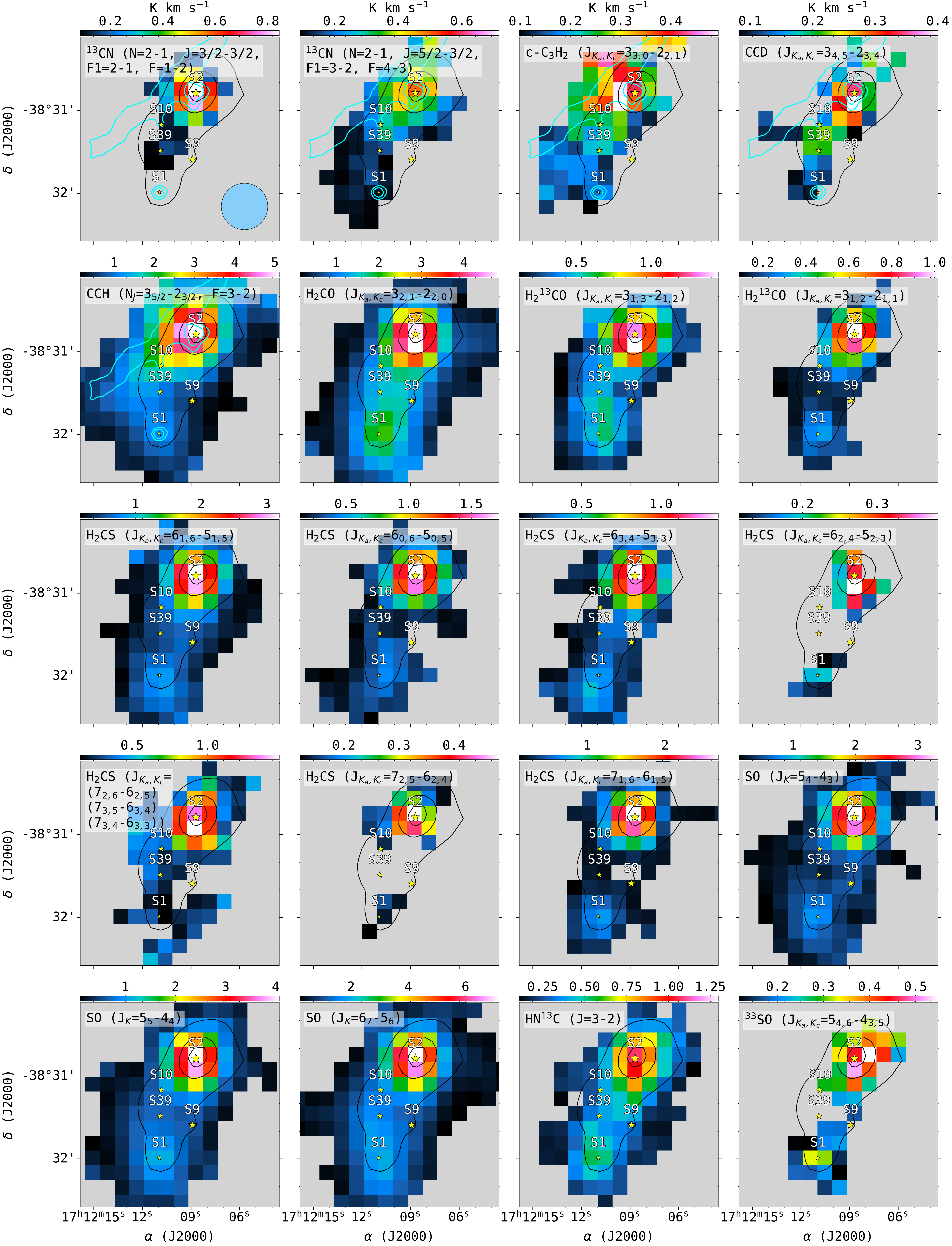}
% \figplaceholder{0.95\textwidth}{120mm}
\caption{Integrated intensity (0-th moment) maps of the molecular
emission. Young stellar objects (YSOs) are shown as yellow stars, the
marker size being proportional to the mass of each YSO. The black
contours show the dust emission at 870\,$\mu$m, at levels of 2.0, 6.0
and 10.0\,Jy\,beam$^{-1}$; the blue contours show the emission at
70\,$\mu$m, at levels of 0.47 and 1.0\,Jy\,pixel$^{-1}$. The blue
circle in the upper left corner of the $^{13}$CN panel represents the
degraded APEX beam size.}
\label{fig:moment0_p1}
\end{figure}

\begin{figure}[p]
\centering
\includegraphics[width=\textwidth]{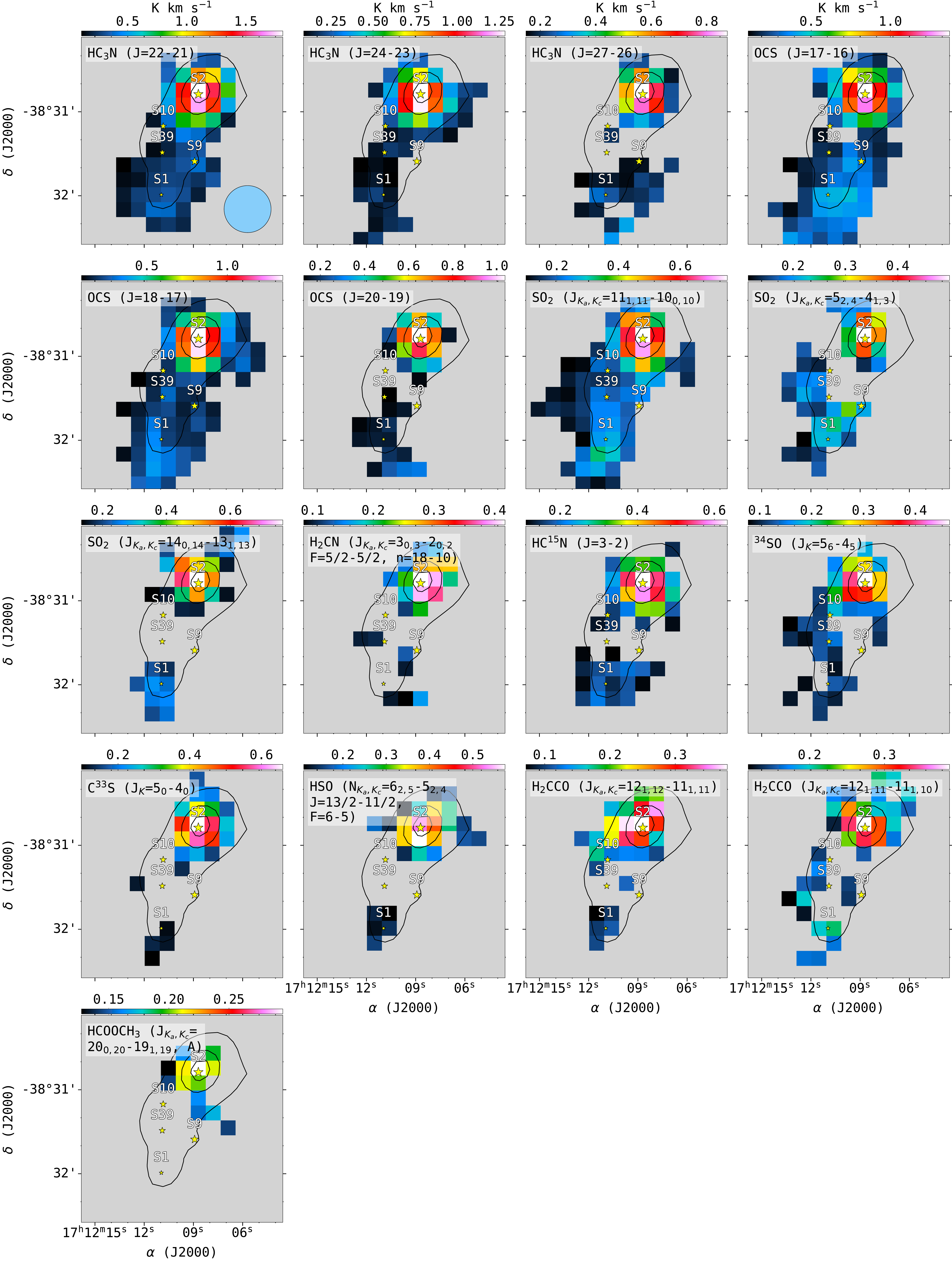}
% \figplaceholder{0.95\textwidth}{120mm}
\caption{Cont.}
\label{fig:moment0_p2}
\end{figure}

The molecules of the first group emit primarily along the edge of the
ionized region. These are the diatomic molecule $^{13}$CN and the small
hydrocarbons c-C$_3$H$_2$, CCD and CCH, whose formation is associated
with the penetration of UV radiation into the molecular gas.

The second group includes molecules whose emission spatially coincides
with the 870\,$\mu$m dust continuum emission, that is, with the entire
region containing the observed YSOs. This group includes both diatomic
molecules and more complex compounds: H$_2$CO, H$_2^{13}$CO, H$_2$CS,
HN$^{13}$C, SO, $^{33}$SO, HC$_3$N, OCS and SO$_2$.

The third group consists of molecules whose emission is observed only
towards YSO S2 and is practically or completely absent towards the
other YSOs. This group includes the nitrogen-bearing molecules
H$_2$CN and HC$^{15}$N, the sulfur-bearing molecules $^{34}$SO,
C$^{33}$S and HSO, and the organic molecules H$_2$CCO and HCOOCH$_3$.%

\subsection{Excitation Temperatures of Rotational Levels}
\label{sec:trot}

\begin{figure}[tbp]
\centering
\begin{minipage}[t]{0.48\textwidth}
  \centering
  \includegraphics[width=\linewidth]{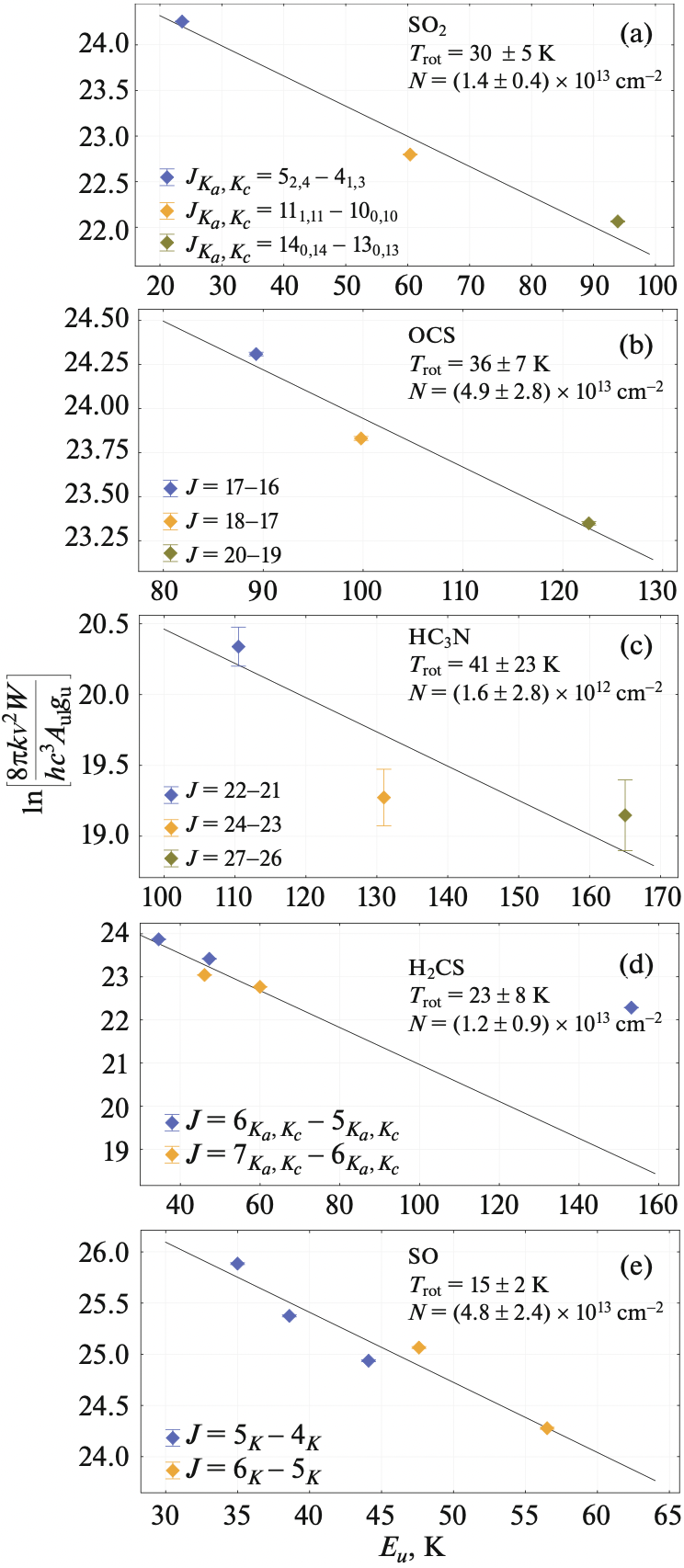}
  \caption{Population diagrams of the rotational levels of SO$_2$, OCS,
  HC$_3$N, H$_2$CS and SO in YSO~S1.}
  \label{fig:popS1}
\end{minipage}\hfill
\begin{minipage}[t]{0.48\textwidth}
  \centering
  \includegraphics[width=\linewidth]{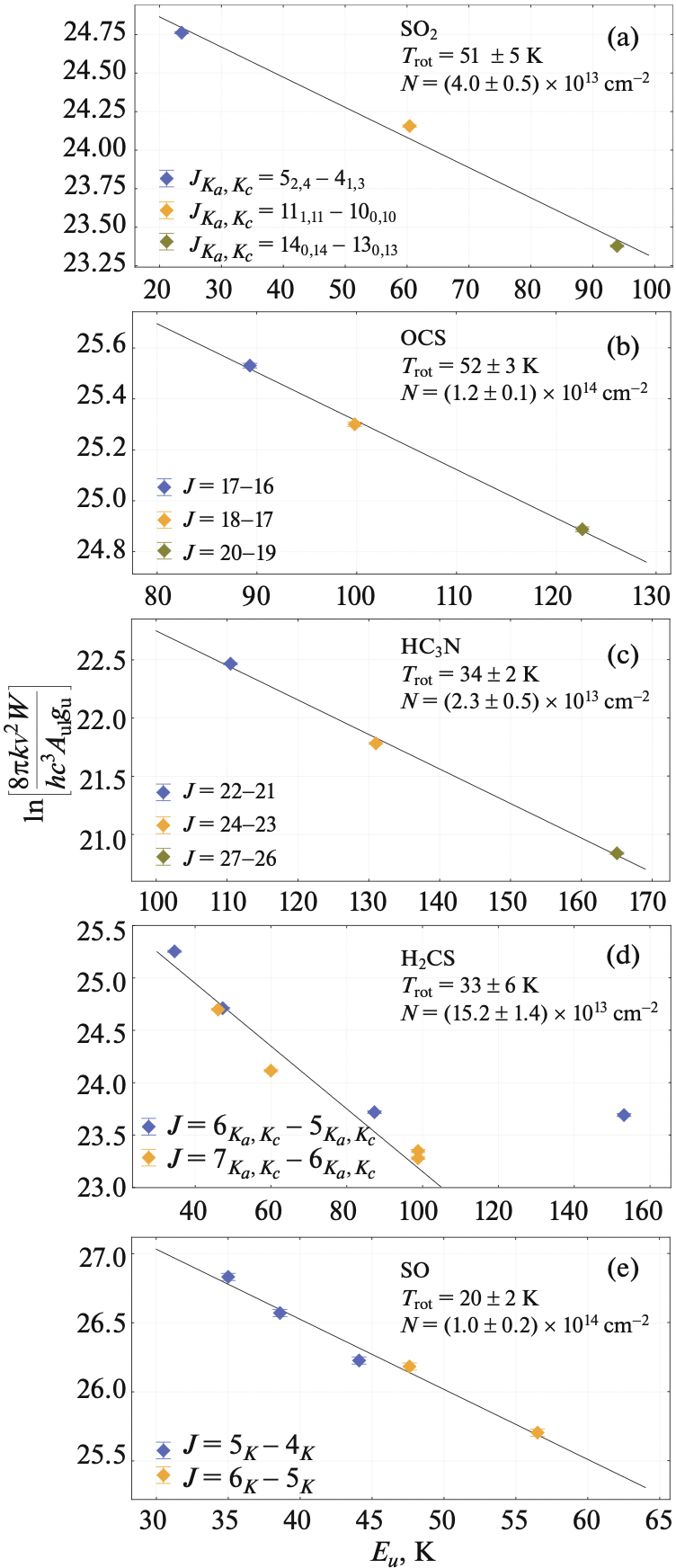}
  \caption{Population diagrams of the rotational levels of SO$_2$, OCS,
  HC$_3$N, H$_2$CS and SO in YSO~S2.}
  \label{fig:popS2}
\end{minipage}
\end{figure}

\begin{figure}[tbp]
\centering
\includegraphics[width=0.5\textwidth]{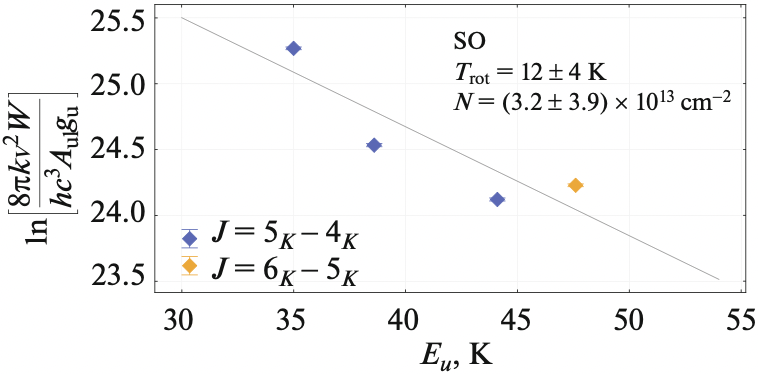}
% \figplaceholder{0.62\textwidth}{50mm}
\caption{Population diagram of the rotational levels of SO in YSO S9.}
\label{fig:popS9}
\end{figure}

\begin{figure}[tbp]
\centering
\begin{minipage}[t]{0.48\textwidth}
  \centering
  \includegraphics[width=\textwidth]{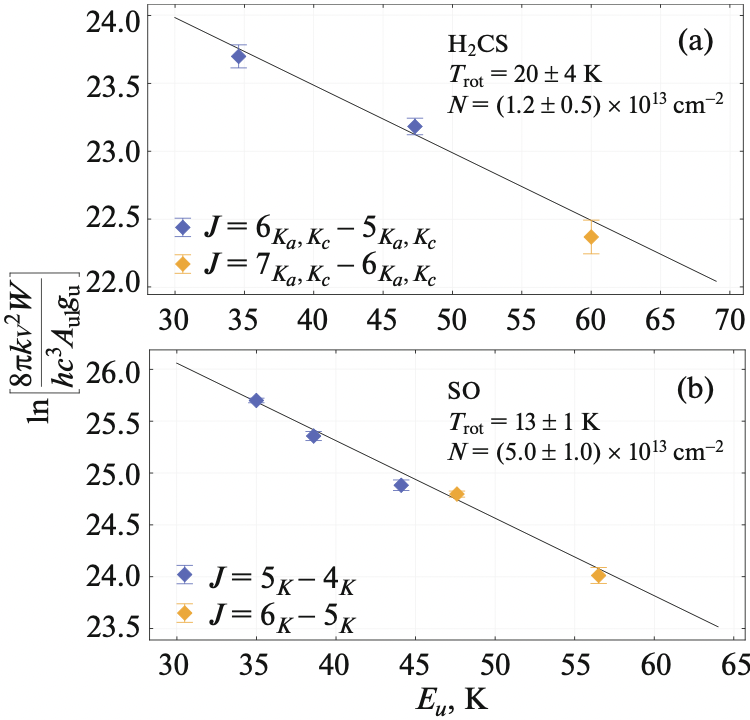}
% \figplaceholder{0.62\textwidth}{60mm}
\caption{Population diagrams of the rotational levels of SO and
H$_2$CS in YSO S10.}
\label{fig:popS10}
\end{minipage}\hfill
\begin{minipage}[t]{0.48\textwidth}
  \centering
  \includegraphics[width=\textwidth]{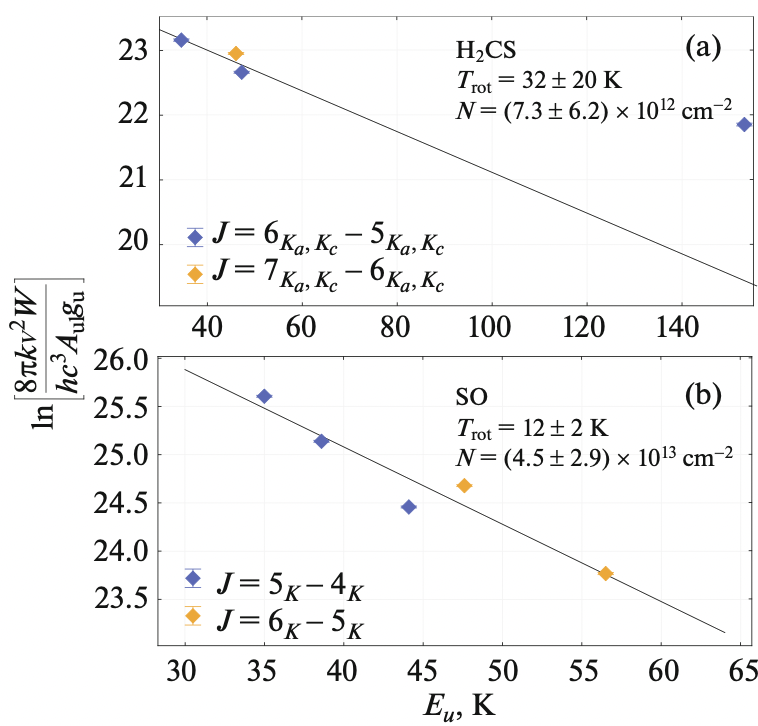}
% \figplaceholder{0.62\textwidth}{60mm}
\caption{Population diagrams of the rotational levels of SO and
H$_2$CS in YSO S39.}
\label{fig:popS39}
\end{minipage}
\end{figure}

The population diagrams of the rotational levels of the molecules
towards the YSOs are shown in Figs.~\ref{fig:popS1}--\ref{fig:popS39}.
The largest number of diagrams---five---was constructed for YSOs S2 and
S1, for the molecules SO$_2$, OCS, HC$_3$N, H$_2$CS and SO. The
signal-to-noise ratio for the other YSOs is lower, so fewer diagrams
could be constructed for them. The figures show that the population
diagrams of SO, SO$_2$, OCS and HC$_3$N follow a linear dependence
between the upper level energy $E_u$ and the logarithm of the column
density. The excitation conditions of the molecules in the YSOs are
therefore consistent with the temperatures derived from the population
diagrams. An exception is the set of diagrams for H$_2$CS, in which
only the transitions with $E_u < 90$\,K lie on a single straight line.
Transitions with higher $E_u$ were not used when constructing the
population diagrams, although they are shown in the figures.

% \begin{figure}[tbp]
% \centering
% \includegraphics[width=0.5\textwidth]{figs/RD_S1.png}
% % \figplaceholder{0.62\textwidth}{90mm}
% \caption{Population diagrams of the rotational levels of SO$_2$, OCS,
% HC$_3$N, H$_2$CS and SO in YSO S1.}
% \label{fig:popS1}
% \end{figure}

% \begin{figure}[tbp]
% \centering
% \includegraphics[width=0.5\textwidth]{figs/RD_S2.png}
% % \figplaceholder{0.62\textwidth}{90mm}
% \caption{Population diagrams of the rotational levels of SO$_2$, OCS,
% HC$_3$N, H$_2$CS and SO in YSO S2.}
% \label{fig:popS2}
% \end{figure}

The highest rotational temperatures $\Trot$ for YSO S2---51 and
52\,K---were obtained from the population diagrams of SO$_2$ and OCS,
respectively, and the lowest value, 20\,K, was found from the
population diagram of SO. In order of decreasing rotational
temperature, the molecules line up in the sequence OCS, SO$_2$,
HC$_3$N, H$_2$CS, SO, which may be related to the chemical structure of
YSO S2.

For YSO S1 the highest rotational temperature, 41\,K, was obtained for
HC$_3$N, but this value is questionable because the spectrum is noisier
for some of the transitions. With the exception of HC$_3$N, the values
obtained for S1 are on average lower than those for S2 by about
5--20\,K, with the smallest difference for SO and the largest for
SO$_2$. It is worth noting that in S1 the molecules (except for
HC$_3$N) are arranged in the same order of decreasing rotational
temperature as in YSO S2.

Towards YSOs S10 and S39, a sufficient number of lines for constructing
population diagrams was observed only for SO and H$_2$CS. The
rotational temperature of S10 determined from the H$_2$CS transitions
is 20\,K, significantly higher than the 13\,K derived from the SO
transitions (Fig.~\ref{fig:popS10}). For S39 it is estimated at 32\,K
from the H$_2$CS transitions and 12\,K from the SO transitions
(Fig.~\ref{fig:popS39}). For S9, only the excitation temperature of SO
could be determined, and it is 12\,K (Fig.~\ref{fig:popS9}).

% \begin{figure}[tbp]
% \centering
% \includegraphics[width=0.5\textwidth]{figs/RD_S10.png}
% % \figplaceholder{0.62\textwidth}{60mm}
% \caption{Population diagrams of the rotational levels of SO and
% H$_2$CS in YSO S10.}
% \label{fig:popS10}
% \end{figure}

% \begin{figure}[tbp]
% \centering
% \includegraphics[width=0.5\textwidth]{figs/RD_S39.png}
% % \figplaceholder{0.62\textwidth}{60mm}
% \caption{Population diagrams of the rotational levels of SO and
% H$_2$CS in YSO S39.}
% \label{fig:popS39}
% \end{figure}

The general conclusion to be drawn from the population diagrams is that the
emission of a given molecule in different YSOs arises under similar
conditions; in YSO S2 the lines are brighter, apparently because of the
higher molecular column density.

\subsection{Abundances of Molecules}
\label{sec:abundances}

\begin{figure}[H]
\centering
\includegraphics[width=0.8\textwidth]{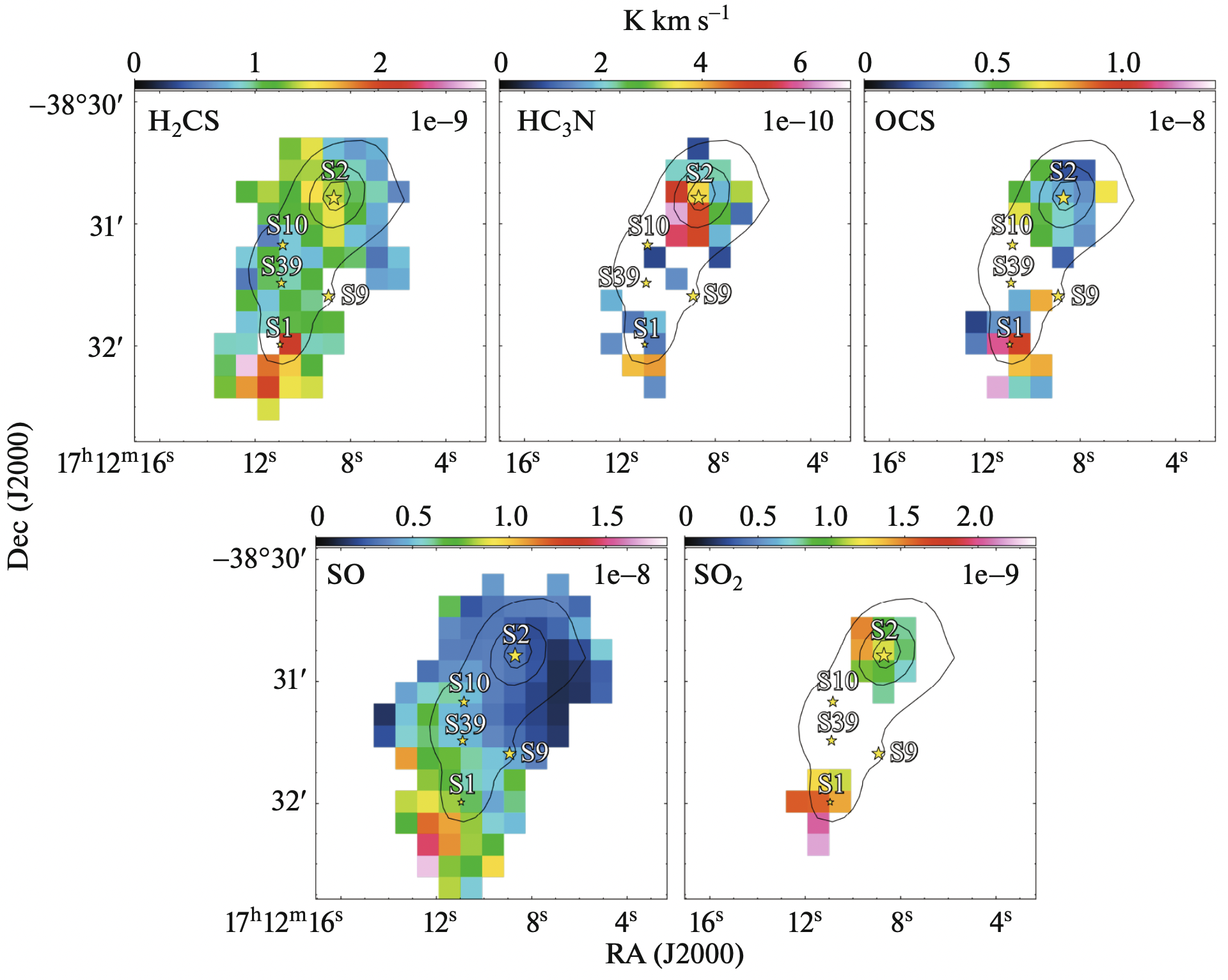}
% \figplaceholder{0.95\textwidth}{100mm}
\caption{Abundances of the molecules. YSOs are shown as yellow stars,
the marker size being proportional to the mass of each YSO. The black
contours show the dust emission at 870\,$\mu$m, at levels of 2.0, 6.0
and 10.0\,Jy\,beam$^{-1}$.}
\label{fig:abund}
\end{figure}

Using the $\NH$ distribution map, we estimated the relative abundances
of H$_2$CS, HC$_3$N, OCS, SO and SO$_2$ in the observed region (in
addition to the molecules considered in the previous paper). The
corresponding maps are shown in Fig.~\ref{fig:abund}, and the abundance
values towards the YSOs are given in Table~\ref{tab:abund}.
\vspace{-0.5cm}

\begin{table}[H]
\centering
\caption{Abundances of SO$_2$, OCS, HC$_3$N, H$_2$CS and SO towards the
YSOs in RCW\,120. The molecular abundances $X$ are given as the ratio
of the molecular column density to the hydrogen nucleus column density
$\NH$.}
\label{tab:abund}
\begin{tabular}{cccccc}
\toprule
& \multicolumn{5}{c}{Abundance} \\
\cmidrule(l){2-6}
YSO & $X(\mathrm{SO_2})$ & $X(\mathrm{OCS})$ & $X(\mathrm{HC_3N})$ &
      $X(\mathrm{H_2CS})$ & $X(\mathrm{SO})$ \\
\midrule
S1  & $1.40 \times 10^{-9}$ & $9.65 \times 10^{-9}$ & $1.27 \times 10^{-10}$ & $2.16 \times 10^{-9}$   & $6.43 \times 10^{-9}$ \\
S2  & $1.14 \times 10^{-9}$ & $3.41 \times 10^{-9}$ & $3.70 \times 10^{-10}$ & $1.33 \times 10^{-9}$   & $3.11 \times 10^{-9}$ \\
S9  & --                    & --                    & --                     & --                      & $4.56 \times 10^{-9}$ \\
S10 & --                    & --                    & --                     & $8.32 \times 10^{-10}$  & $4.95 \times 10^{-9}$ \\
S39 & --                    & --                    & --                     & $9.34 \times 10^{-10}$  & $4.86 \times 10^{-9}$ \\
\bottomrule
\end{tabular}
\end{table}

The abundances of the sulfur-bearing species in YSO S1 are higher than in
YSO S2, the largest excesses being factors of two and three for SO and
OCS, respectively. The abundance of HC$_3$N, in contrast, is twice as
high in YSO S2 as in YSO S1. Because of the low brightness of the
lines, a reliable abundance estimate for the remaining YSOs could be
obtained only for SO and H$_2$CS.%

\subsection{Correlations of the Molecular Column Densities}
\label{sec:corrres}

Combining the column density distributions from this work with the
results of \citet{Plakitina2024}, we examined the correlations of the
column densities $N$ of 20 different molecules in the observed region
and determined the Pearson correlation coefficient $p$ for them. In
addition, we analysed the correlation of the $N$ values of the observed
molecules with the column density of hydrogen, $\NH$, also taken from
the previous work. The results are shown in Fig.~\ref{fig:corrmatrix}.

A positive linear correlation is observed for most pairs of molecules.
Values $p > 0.95$ were obtained for the pairs CH$_3$CCH--CH$_3$CN,
CH$_3$CCH--DCN, CH$_3$CCH--H$^{13}$CN, CH$_3$CN--SO,
$^{13}$CO--C$^{18}$O, DCN--SO, DCN--H$^{13}$CN, H$^{13}$CN--SO,
H$^{13}$CN--H$^{13}$CO$^{+}$ and H$^{13}$CO$^{+}$--H$_2$CO.%

High values of $p$ for a pair of molecules indicate that they form in a
similar way, namely in gas-phase reactions or on the dust surface.
Correlations in the pairs DCN--H$^{13}$CN, $^{13}$CO--C$^{18}$O,
H$^{13}$CO$^{+}$--DCO$^{+}$ and CS--C$^{34}$S are expected, since these
are pairs of isotopologues. Interestingly, despite the high correlation
coefficient of 0.94, CS and C$^{34}$S were assigned to different
spatial groups---although it should be acknowledged that the division
of molecules into groups 2 and 3 is rather subjective.

It is interesting to note the weak correlation of the column densities of methanol and of the CS isotopologues with the column density of hydrogen. This may indicate that their emission arises in compact regions that are not strongly correlated with the total column density along the line of sight. For the pairs CH$_3$OH--CCH, CH$_3$OH--CH$_3$CCH, CH$_3$OH--CH$_3$CN, $^{13}$CO--CH$_3$OH, $^{13}$CO--CS, C$^{18}$O--CS and H$_2$CO--CS the Pearson correlation coefficients are also close to zero, which suggests that the formation of these molecules proceeds independently. The fact that some pairs of molecules show negative correlations formally indicates that the formation of one molecule of the pair leads to the destruction of the other or, in a broader sense, that the physical conditions favourable for the formation of one molecule are unfavourable for the formation of the other.

\begin{figure}[H]
\centering
\begin{overpic}[width=\textwidth]{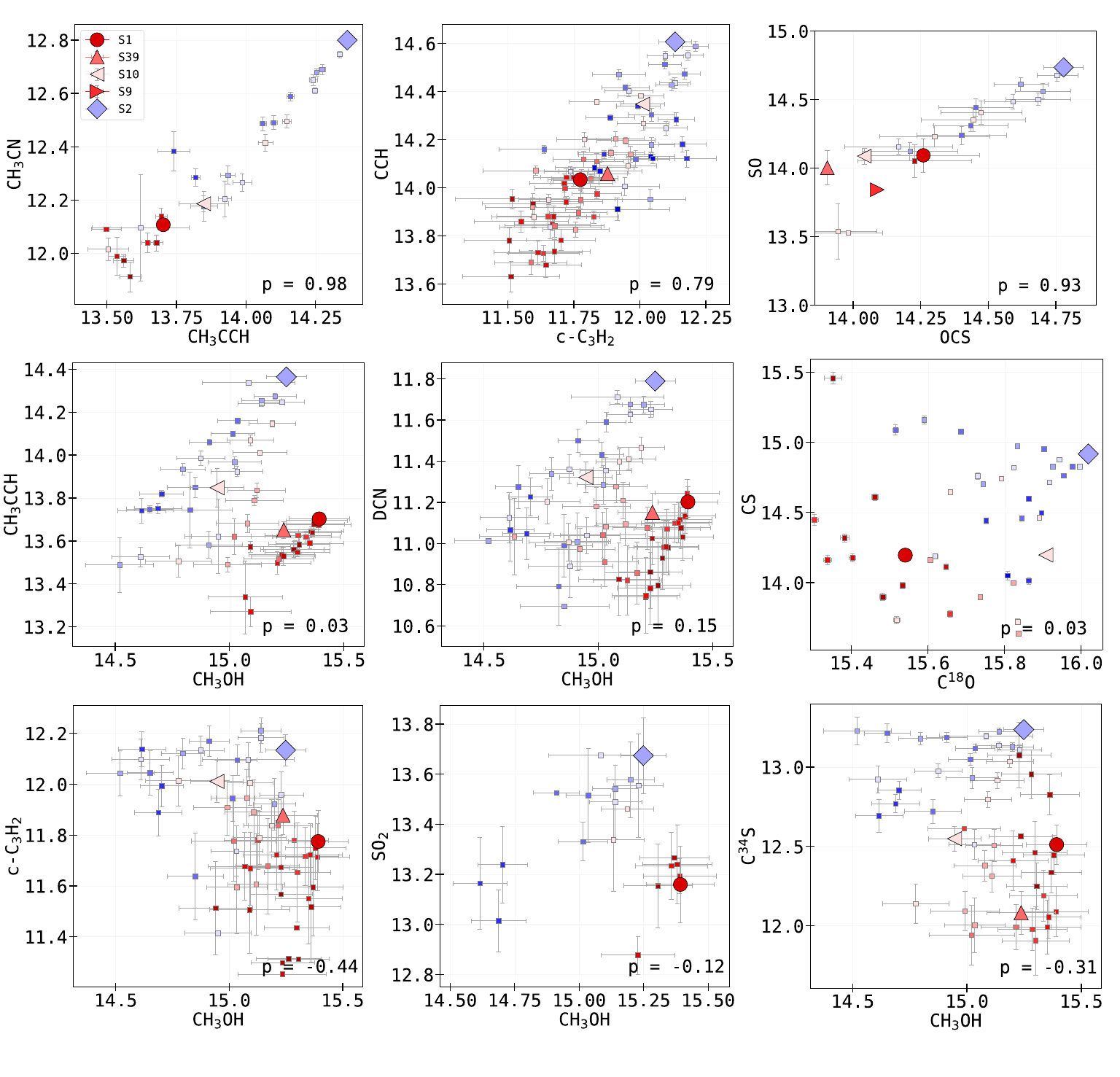}
  \put(20,70){\small (a)}  \put(52,70){\small (b)}  \put(86,70){\small (c)}
  \put(20,39.5){\small (d)}  \put(52,39.5){\small (e)}  \put(86,39.5){\small (f)}
  \put(20, 9){\small (g)}  \put(52,9){\small (h)}  \put(86, 9){\small (i)}
\end{overpic}
% \includegraphics[width=\textwidth]{figs/CorPlots.pdf}
% \figplaceholder{0.95\textwidth}{110mm}
\caption{Pixel-by-pixel comparison of the molecular column densities.
The more saturated the red or blue hue, the further south or north the
pixel lies. Large symbols mark the pixels corresponding to the five
YSOs (see legend). The Pearson correlation coefficient is given in each
panel.}
\label{fig:pixelcorr}
\end{figure}
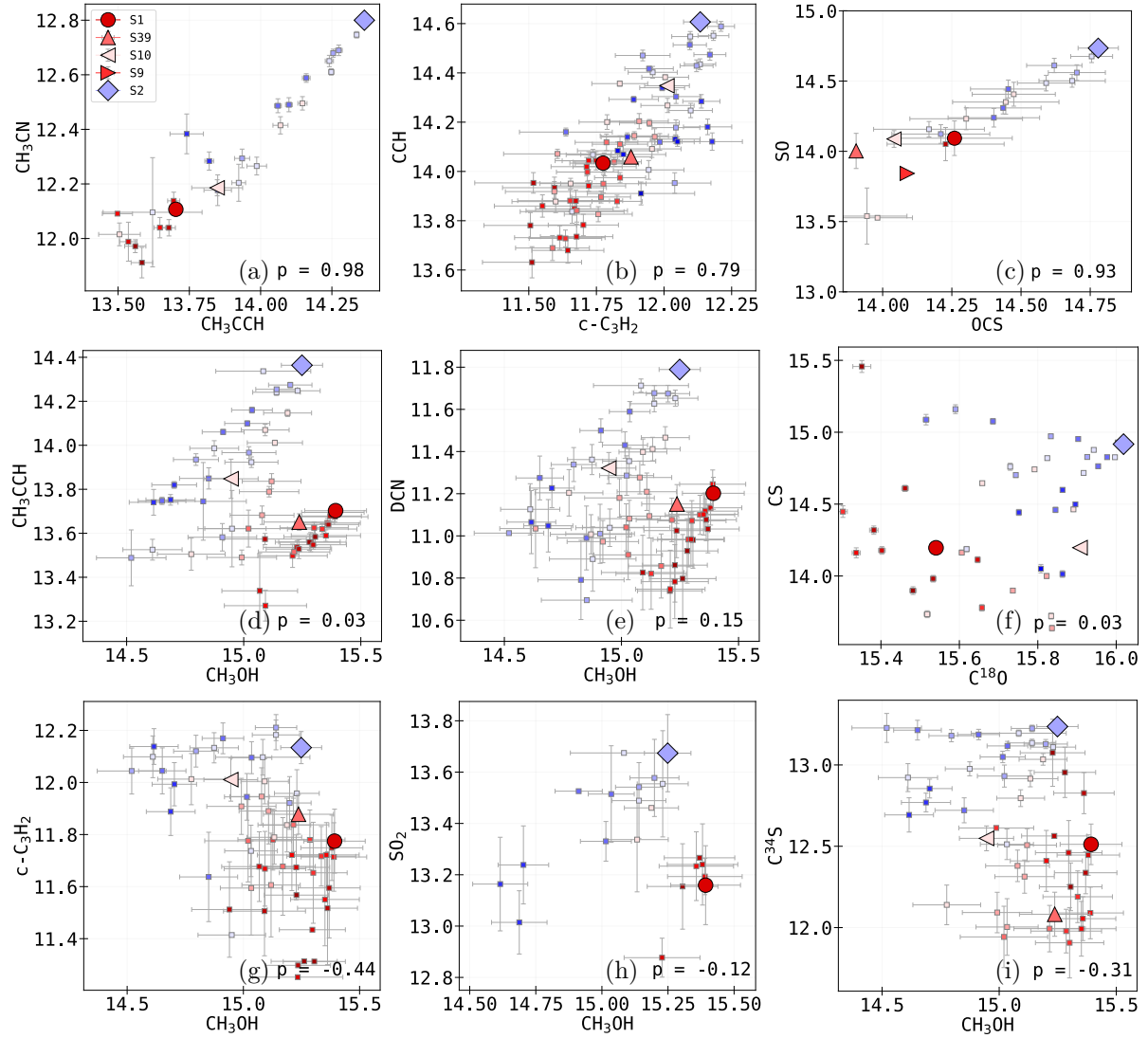

The real situation, however, is somewhat more complicated. For a more detailed study of the correlations, and to verify which parts of the observed region show a strong correlation of the molecular column densities, we performed a pixel-by-pixel comparison of the maps and constructed correlation plots of the column densities for pairs of molecules as a function of position in the observed area (Fig.~\ref{fig:pixelcorr}). Each pixel in the plots is coloured according to a colour scale from red to blue, which shows the distance from the declination $\delta = -38^{\circ}31'11''$. Blue pixels correspond to the northern part of the map, in which YSO S2 is located, and red pixels to the southern part, containing YSO S1. The more saturated the shade of red or blue, the further south or north the pixel lies. Note that with the chosen division into northern and southern parts, only YSO S2 lies in the northern part and all the other YSOs lie in the southern part.

Panels (a)--(c) of Fig.~\ref{fig:pixelcorr} show pairs of molecules with $p \geq 0.75$. Here the pixels of the southern and northern parts of the maps fall on a common correlation: the southern part of the map contributes to the lower part of the plot and the northern part to the upper part. Such pairs include, for example, CH$_3$CN and CH$_3$CCH, CCH and c-C$_3$H$_2$, CCH and CH$_3$CN, SO and H$_2$CS, and SO and OCS. Panels (d)--(f) show pairs of molecules for which the Pearson correlation coefficient is close to zero. The plots make clear that we are in fact sometimes dealing not with the absence of a correlation but with different correlations in the northern and southern parts of the region. If the ``northern'' and ``southern'' pixels in Figs.~\ref{fig:pixelcorr}d--f are considered separately, two positive correlations are seen, but with different abundance values. In particular, the methanol abundance is higher in the southern part of the region than in the northern part. Even where there is no visible correlation, as in the C$^{18}$O--CS pair (Fig.~\ref{fig:pixelcorr}f), the different locations of the northern and southern pixels in the diagram are clearly visible.

In the pairs c-C$_3$H$_2$--CH$_3$OH, SO$_2$--CH$_3$OH and
C$^{34}$S--CH$_3$OH, presented in panels (g)--(i) of
Fig.~\ref{fig:pixelcorr}, the correlation is negative, but in this case
we are most likely again dealing with different correlations in the
northern and southern pixels. In general, for most of the pairs with
small and negative $p$ values we are dealing with a global
anticorrelation between the northern and southern regions, combined
with a positive correlation within each region.

Interestingly, for some pairs of molecules a high correlation
coefficient between the column densities can be predicted from the
appearance of their integrated intensity maps. This is the behaviour of,
for example, the photodissociation-region tracers CCH and
c-C$_3$H$_2$, as well as of the molecules whose bright emission is
observed only towards YSO S2, such as H$_2$CS and SO, or CH$_3$CN and
CH$_3$CCH (the integrated intensity maps of these two pairs are shown
in Fig.~3 of \citealt{Plakitina2024}). This is not the case for the
CCH--CH$_3$CN pair, however: despite the large value of $p$, the
integrated intensity distributions of these molecules differ on the
maps. In other words, a significant correlation of column densities
does not necessarily imply a similarity in the spatial distribution of
the integrated intensity, but may instead reflect a similarity in the
pathways of molecule formation in the observed regions, as discussed
below.

To identify possible causes of the elevated methanol abundance in the
southern part of the cloud, we modelled the object with the
\code{Presta} astrochemical model. Varying the model parameters within
the ranges given in Section~\ref{sec:model}, we found that the typical
abundances of CH$_3$CCH, CH$_3$CN and CH$_3$OH observed in the object
are best reproduced, to within an order of magnitude, at a model time
$t_{\mathrm{model}} = 10^{5}$\,yr in a model with hydrogen number
density $n_{\rm H} = 10^{4}\,\percc$ and visual extinction
$\Av \sim 3\mg-5\mg$.%
\begin{figure}[H]
\centering
% \begin{minipage}[t]{0.48\textwidth}
  \centering
\includegraphics[width=0.5\textwidth]{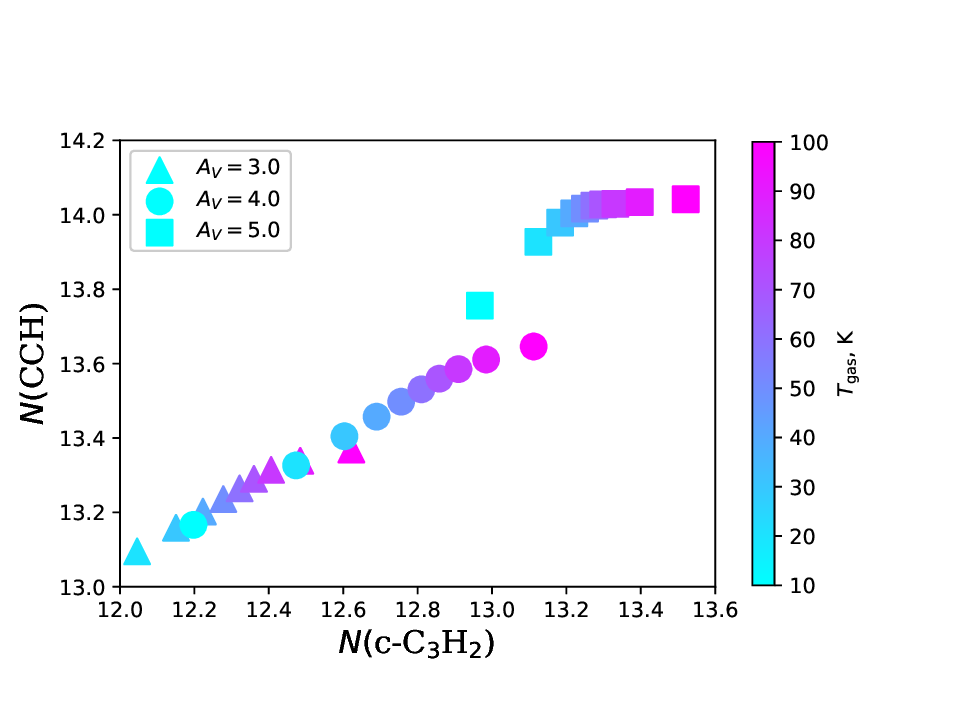}\\[-0.5cm]
\includegraphics[width=0.5\textwidth,trim=0 0 0 10mm,clip]{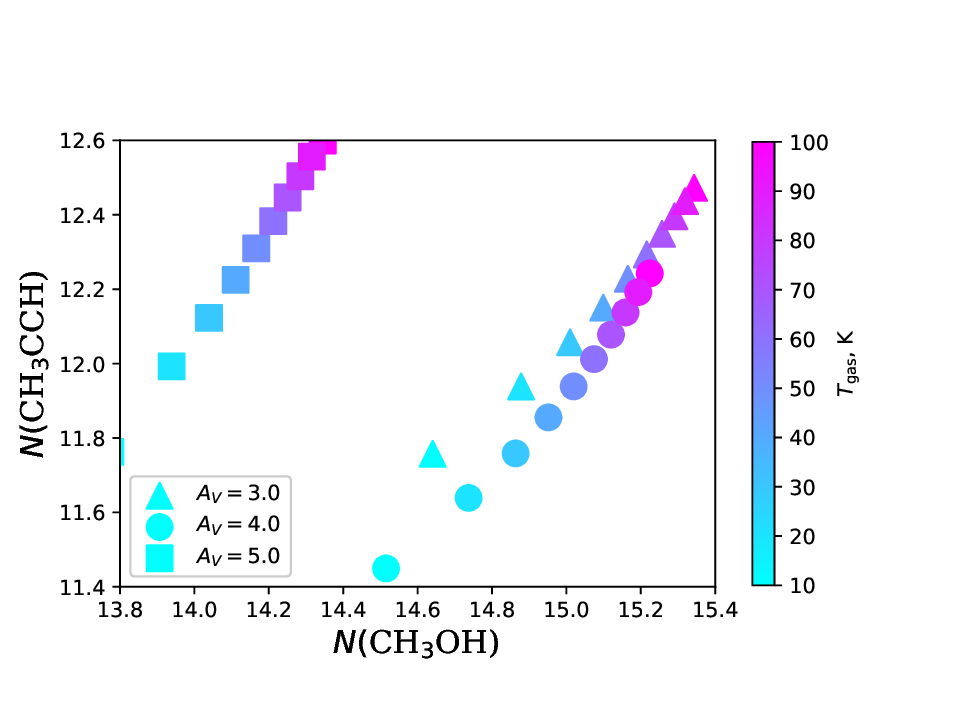}
% \figplaceholder{0.95\textwidth}{70mm}
\caption{Simulation results for the best model at a time of $10^{5}$
years. The model calculations are shown as triangles for the model with
$\Av = 3\mg$, circles for $\Av = 4\mg$ and squares for $\Av = 5\mg$.
The colour scale of the markers corresponds to different values of
$\Tgas$.}
\label{fig:model}
% \end{minipage}
\end{figure}

 Figure~\ref{fig:model} show the model
abundances of the pairs CCH--c-C$_3$H$_2$ and CH$_3$CCH--CH$_3$OH at different gas temperatures and for $\Av = 3\mg$, $4\mg$ and $5\mg$; a clear difference is seen between the abundance sequences of CH$_3$CCH and CH$_3$OH.

These sequences are consistent with the division into two branches
observed in Fig.~\ref{fig:pixelcorr}, where, at comparable CH$_3$CCH
abundances, the methanol abundance in the southern part of the cloud
exceeds that in the northern part by up to an order of magnitude. Such
a theoretical regularity may be related to the fact that CH$_3$CCH
forms in the gas phase whereas CH$_3$OH forms on the dust surface. In a
medium with $\Av \leq 4\mg$, photodesorption is more efficient than in
a medium with $\Av = 4\mg$, and the methanol abundance in the gas phase becomes higher. A further
decrease in the extinction to $\Av \leq 2\mg$ leads to methanol being
effectively destroyed by photodissociation after desorption, so that
its gas-phase abundance decreases. We therefore conclude that the
increased methanol abundance in the southern part of the molecular
cloud may be due to photodesorption associated with a slightly lower
average value of $\Av$. Comparing the distribution of $\NH$ in the
vicinity of YSOs S1 and S2 \citep[see][]{Plakitina2024}, we see that
the difference between the hydrogen column densities of these objects
is greater than $1.81 \times 10^{21}\,\persc$ (equivalent to
$\Av = 1\mg$), so our assumption of a lower $\Av$ near YSO S1 is
justified. Photodissociation also affects molecules such as CH$_3$CCH,
but their abundance is largely restored by gas-phase reactions.

%======================================================================
\section{DISCUSSION OF RESULTS}
\label{sec:discussion}

\subsection{Signs of Hot Gas in the YSOs}
\label{sec:hotgas}

Our results indicate that the observed molecules form under different
conditions and, accordingly, trace different stages of the star
formation process. The question arises of how far our estimated
abundances can indicate the presence of gas heated by protostars in the
YSOs studied.

To assess this possibility, we compared the molecular abundances
towards the YSOs obtained in this paper and in \citet{Plakitina2024}
with the results of chemodynamical modelling of the collapse of a
protostar \citep{Kochina2025}, in which the same initial abundances
were used as in our study. A comparison of the spatial scale of the
model collapsing cloud with the maps in Fig.~\ref{fig:abund} shows that
the entire computational domain falls within a single pixel of our
maps. The model allows the moment of protostar formation to be traced,
which leads to the heating of gas and dust in its immediate vicinity.
The heating is accompanied by an increase in the molecular abundances
near the inner boundary of the computational domain. The heated region
is small, however, 100--200\,AU from the protostar. The abundances we
obtained therefore reflect the abundances of molecules in the envelope
of YSO S2, whose size is up to 40\,000\,AU, and do not reflect the
increase in molecular abundances in the immediate vicinity of the
protostar.

The only evidence of the presence of hot gas towards YSO S2, and
possibly also towards YSOs S1 and S39, is the H$_2$CS
($6_{3,4}$--$5_{3,3}$) transition, with an upper level energy
$E_u = 153.1$\,K (Table~\ref{tab:spectro}). Its observed intensity far
exceeds that expected for the rotational temperatures of 20--30\,K
derived from the lower-energy transitions.

\citet{Kirsanova2021} detected lines with $K \geq 4$ in the CH$_3$CN
($12_K$--$11_K$) series towards YSO S2. Those authors concluded that
these lines appear because of the presence not only of an extended
molecular cloud along the line of sight, but also of a compact hot core
at an early stage of formation. Our detection of a relatively bright
H$_2$CS ($6_{3,4}$--$5_{3,3}$) line is consistent with that result. The
presence of compact emission in the CH$_3$CN lines of YSO S2 was first
found by \citet{Figueira2018}, although they did not analyse the
physical conditions in the molecular gas. The presence of a compact hot
core in YSO S2 is also supported by the detection of the methyl formate
line HCOOCH$_3$ ($20_{0,20}$--$19_{1,19}$), see
Table~\ref{tab:spectro}. As shown by \citet{Kochina2024}, methyl
formate forms on dust at the dark cloud stage and appears in the gas
phase during the formation of a protostar in a molecular cloud, at the
hot core stage. This line was detected only in YSO S2, owing to the
heating of dust in a compact source that is not present towards the
other YSOs.

\subsection{Peculiarities of the Destruction of Dust Grain Mantles
            Near the YSOs}
\label{sec:mantles}

It was suggested above that the increased methanol abundance in the
southern part of the object may be due to more efficient
photodesorption. Considering other types of desorption, we note that
the thermal desorption rate, which depends on $\Tdust$, should be
comparable in YSOs S1 and S2, since the dust temperatures of these
objects agree to within 1\,K \citep{Plakitina2024}. Moreover, $\Tdust$
in these YSOs is lower than the sublimation temperature of methanol
molecules, which is about 80\,K \citep[e.g.,][]{Wiebe2019}. The
desorption rate due to cosmic rays should also be the same, since these
YSOs lie within the same molecular clump at the boundary of the \HII{}
region, whose size is about 1\,pc. An important factor that is not
taken into account in the present modelling is the shock waves and
high-velocity outflows that accompany the birth of stars
\citep{Beuther2005,Arce2007,Zinchenko2015}. The presence of outflows in
the vicinity of YSO S1 is indicated by the broad wings of the methanol
lines \citep{Kirsanova2021,Plakitina2024} and by the detection of
class~I methanol maser emission \citep{Voronkov2014}, which forms in
regions covered by outflows \citep{Voronkov2012}. Shock waves are not
explicitly incorporated into the \code{Presta} model, but their effect
could be accounted for by introducing an additional desorption process.%

 By including such a mechanism at a qualitative level in our model, we
could obtain a result similar to that shown in
Fig.~\ref{fig:model}. We therefore cannot exclude a contribution of the
destruction of dust grain mantles during grain collisions to the
enhancement of the methanol abundance in the gas phase.

\citet{Oberg2021}, summarizing the results of astrochemical modelling
and of observations of star-forming regions, showed that the structure
of the ice mantles of dust grains is heterogeneous: the
water-dominated ice layer is covered by a layer rich in molecules such
as CO, CO$_2$ and methanol. Moreover, the water layer forms earlier
than the CO layer, since the latter forms at a late stage in the
development of a dense cold molecular cloud, when CO begins to freeze
out of the gas onto the dust. The formation of abundant methanol in the
upper part of the mantle is associated with the sequential
hydrogenation of CO in reactions with atomic hydrogen
\citep{Watanabe2002,Fuchs2009,Punanova2022}. Having analysed
observations of the CO$_2$, OCS and methanol lines towards massive
protostars, \citet{Santos2024} showed that SO, CO$_2$ and OCS form
earlier than methanol, at the stage of molecular cloud development when
the ice is predominantly water ice. Methanol forms on the dust grains
later, mainly in the upper CO layer.

Our results are consistent with the presence of water and CO layers in
the dust grain mantles. Figure~\ref{fig:pixelcorr}c shows a high
($p = 0.93$) correlation between the SO and OCS column densities; the southern part of the cloud does not stand out for an enhanced
abundance of either of these molecules. In contrast, there is no correlation between methanol and SO$_2$ ($p = -0.12$, see
Fig.~\ref{fig:corrmatrix}), even though both molecules form on dust
grains. The correlation between OCS and methanol, although positive, is
weak ($p = 0.33$). This can be explained if we take into account that
the outer mantle layers, rich in methanol, are destroyed first, which
is consistent with three-phase astrochemical models in which chemical
reactions occur mainly in the near-surface layers
\citep{Borshcheva2022}.

Summarizing the above discussion of the correlation plots, we conclude
that the enhanced abundance of gas-phase methanol in the southern part
of the cloud is caused by the destruction of the upper layers of the
dust mantles by photodesorption or, possibly, by shock waves. This
process affects to a lesser extent the abundances of molecules that
form in the gas phase (for example, CH$_3$CCH and CH$_3$CN). For this
reason the correlation plots display fork-shaped patterns
(Fig.~\ref{fig:pixelcorr}).

%======================================================================
\section{CONCLUSIONS}
\label{sec:conclusions}

We have investigated the chemical composition and the physical
conditions in a dense molecular clump located at the edge of the \HII{}
region of RCW\,120. The clump combines a variety of physical conditions
in a compact volume---from hot ionized gas to cold molecular gas with
embedded young stellar objects. Using APEX observations in the
200--260\,GHz range, we detected 65 lines of 35 molecules towards the
massive YSO S2. The most complex molecule detected was methyl formate,
HCOOCH$_3$, which consists of eight atoms. From the derived molecular
column densities we found that the methanol abundance is higher in the
southern part of the dense clump than in its northern part, while the
abundances of other complex molecules, such as CH$_3$CN and CH$_3$CCH,
are comparable. In addition, the methanol abundance is enhanced
relative to that of other oxygen-bearing molecules, such as SO and OCS.

To interpret the observational results we used the astrochemical model
\code{Presta}. The model reproduces the observed abundances of
CH$_3$CCH and CH$_3$OH and shows that a possible mechanism responsible
for the enhanced gas-phase methanol abundance is photodesorption from
the grain mantles. As the visual extinction $\Av$ decreases from $5\mg$
to $3\mg$, methanol desorption accelerates, while the molecules
released into the gas phase survive, avoiding destruction by UV
radiation, which becomes significant only at $\Av \sim 2\mg$.

The strong linear correlation between the column densities indicates
that the molecules form in the same phase---either in the gas phase or
on dust grains. No correlation is observed when the molecules form in
different phases---one in the gas phase and the other in dust mantles
(e.g., CCH and CH$_3$OH). The weak correlation between methanol and the
other oxygen-bearing molecules (SO, OCS, SO$_2$), which also form on
dust, suggests that in the southern part of the clump only the upper,
CO-rich layers of the dust mantles are destroyed.

%======================================================================
\section*{ACKNOWLEDGMENTS}

The authors thank A.~O.~H.~Olofsson, who carried out the observations
under project ID 0108.F-9313(A).

\section*{DATA AVAILABILITY}

The observational data are available in the ESO archive
(\url{http://archive.eso.org/wdb/wdb/eso/apex/form}), programme ID
0108.F-9313(A).

\section*{FUNDING}

This work was supported by the Russian Science Foundation, grant
No.~24-22-00097.

\section*{CONFLICT OF INTEREST}

The authors declare that they have no conflicts of interest.

%======================================================================
%                            REFERENCES
%======================================================================

%======================================================================
%  APPENDIX -- placed after the bibliography, as is standard for
%  Astrophysical Bulletin / MNRAS / A&A. \appendix switches the section
%  numbering to A, B, ...; the two lines below restart figure numbering
%  so this becomes Fig. A1 rather than continuing the main sequence.
%======================================================================
\appendix
\renewcommand{\thefigure}{A\arabic{figure}}
\renewcommand{\thetable}{A\arabic{table}}
\setcounter{figure}{0}
\setcounter{table}{0}
\newpage
\section{Correlation coefficients of the molecular column densities in RCW\,120}
\label{app:corr}

\begin{figure}[htbp]
\centering
\includegraphics[width=\textwidth]{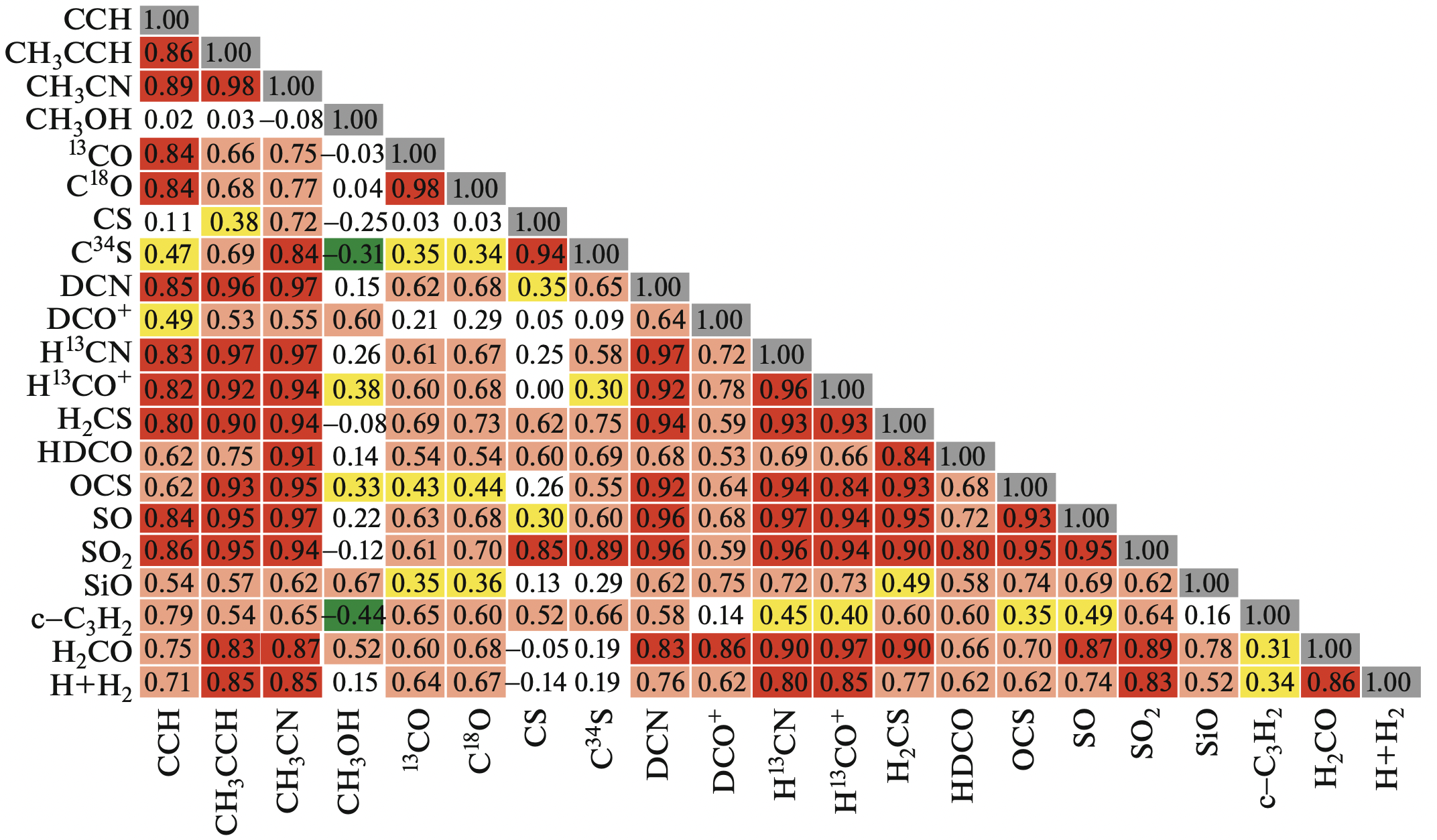}
% \figplaceholder{0.95\textwidth}{110mm}
\caption{Correlation coefficients of the molecular column densities in
RCW\,120. Red cells correspond to Pearson correlation coefficients
$p \geq 0.8$, light brown to $0.5 \leq p < 0.8$, yellow to
$0.3 \leq p < 0.5$, colourless cells to $-0.3 < p < 0.3$ and green to
$-0.5 \leq p \leq -0.3$.}
\label{fig:corrmatrix}
\end{figure}

\end{document}